# Insights into the Limitations of Parameter Transferability in Heteronuclear SAFT-type Equations of State

Emanuel A. Crespo,[a] and João A. P. Coutinho[a*]

[a]CICECO – Aveiro Institute of Materials, Department of Chemistry, University of Aveiro, 3810-193 - Aveiro, Portugal;

*Corresponding author: jcoutinho@ua.pt

## Abstract

The use of heteronuclear models are often viewed as ways to improve the predictive ability and parameter transferability of advanced association models, such as those derived from the Statistical Associating Fluid Theory (SAFT). Indeed, several results in the literature have suggested that this approach can be useful to accurately describe a given family/series of homologous compounds and their mixtures, with accuracies competitive (and in some cases better) than those obtained using the more traditional SAFT variants. However, the parameter transferability of the different groups, i.e. between different families of compounds, without the introduction of new groups or refitting existing ones, is seldom reported, and often overlooked, making an accurate evaluation of the heteronuclear models difficult.

This work analyzes whether the increased complexity of a heteronuclear treatment of a SAFT-type EoS, namely the SAFT-γ-Mie EoS, results in a significant increase on both the predictive ability and parameter transferability of the model, across different families of compounds. This is done by using a case study involving some different (yet related) families of compounds, containing a small number of common functional groups. The selected families include alkanes, alkane-1-ols, glycols, glymes (or glycol ethers), and the non-ionic alkyl polyglycolether ($C_iE_j$) surfactants. These compounds can be deconstructed in only four different functional groups (e.g. $CH_3$, $CH_2$, OH, and $CH_2OCH_2$), only one of which is not found in the widely studied *n*-alkanes and alkan-1-ols series.

The results obtained show that the transferability of group parameters, across different families of compounds, in a heteronuclear SAFT-type EoS does not allow an adequate description of the phase equilibria of these systems. Therefore, to achieve a reasonable accuracy in the description of these systems, a specific refitting of group parameters is required for a given family, or even for a particular system, destroying the predictive capability of these models. Moreover, this increases the number of adjustable parameters to numbers similar to those used in homonuclear approaches, further reducing the advantages of using heteronuclear models.

## 1. Introduction

Thermodynamic models are used in process simulators to estimate phase equilibria and characterize the thermo-physical properties such as transport, thermal, derivative, and interfacial properties, of the systems being studied. They are essential for the accurate design, simulation, control, and economic evaluation of new and existing industrial processes and/or products.[1]

Although, in theory, the required data could be obtained by direct or indirect measurements, both the cost and time required for the task, in particular for multicomponent systems, are prohibitive. Moreover, properties are sometimes needed under extreme ($T$, $P$) conditions that are difficult to replicate, using common experimental setups. Therefore, the prediction and/or correlation of phase equilibrium and thermodynamic properties for pure fluids and mixtures, using accurate and robust thermodynamic models, remains crucial in chemical engineering and related fields.

Traditional approaches include empirical correlations for the description of single-phase properties or the use of Gibbs excess energy ($G^E$) models and cubic equations of state (EoSs) for phase equilibrium calculations. Despite the simplicity and accuracy of activity coefficient models for certain applications, their inability to account for pressure effects (and to describe thermophysical properties), makes the use of EoSs often preferred. Over the last decades, the increasing complexity of the systems of interest to the chemical industry have emphasized the limitations of the well-known and widely applied cubic EoSs (e.g. Peng-Robinson and Soave-Redlich-Kwong). These models, although very accurate for molecules in which the most important intermolecular forces are dispersive, or weak polar interactions, they usually fail when are applied to describe pure fluids and mixtures where strong polar or associating forces play an important role, requiring the development of more sophisticated models, with a more complex physical foundation.[2]

A major breakthrough to improve the accuracy and applicability of EoS models was achieved by applying statistical thermodynamics concepts into the development of advanced molecular-based EoSs, the most promising being those derived from the Statistical Associating Fluid Theory (SAFT), proposed by Chapman et al. in the late 80's.[3–6] Within the framework of classical SAFT variants, molecules are represented as a chainlike fluid where a number of equally-sized spherical monomers are covalently bonded to each other forming chains that may, or may not associate at specific bonding sites to model the presence of hydrogen bonding and other type of strong, short-range and highly directional forces.[6]

Even though SAFT-type EoSs have been around for over 30 years, research on applied thermodynamics, focused on the development and/or improvement of these models, continues to be a very active research topic, with most efforts being devoted to improve one or more of the following characteristics of some of most used SAFT variants:

- Accuracy/Applicability – There is an increasing demand for models that can deal not only with simple, common fluids, but also with complex systems that have recently found widespread application in industry, and for which the performance of the current models is still not optimal. This can include strong associating and polar fluids (biobased solvents, ionic liquids, deep eutectic solvents), or electrolyte systems, to name just a few.
- Extrapolative ability – The amount of experimental data available to parameterize the EoS is often limited in terms of temperature, pressure, and/or composition range. Therefore, thermodynamic models should be able to successfully extrapolate results for thermodynamic conditions, other than those used in the parameterization procedure, without a significant loss of performance.
- Predictive ability – One of the advantages of using EoSs over $G^E$ models is their versatility and potential to provide a wide range of thermodynamic properties. Hence, a good thermodynamic model should be able to provide reliable predictions for properties other than those used in the model's parameterization and to calculate different types of phase equilibrium.
- Parameterization methods – The parameterization method (amount and type of experimental data used in the fitting; weights of the different properties considered, existence of transferable/fixed parameters, …) can have a profound effect on the performance and applicability of any given model. Moreover, alternative parameterization methodologies are often required when the pure compounds are solid, have negligible vapor pressures, and no densities nor critical properties available, hindering the use of traditional parameterization approaches.
- Parameters transferability – Related to the predictive ability of an EoS, there have been reported several attempts to transfer model parameters between different compounds of a given homologous series or different families.

Concerning the latter aspect, several authors have shown that simple empirical relationships between the molecular parameters and the molecular weight/carbon number of the compounds can be drawn for a given homologous series, allowing the inter/extrapolation of the model parameters to compounds of different chain length.[7–11] However, as the compounds become more complex, the performance of the extrapolated parameters deteriorates for increased chain lengths, especially if the underlying coarse-grain model of the smaller oligomers is too simplistic.

To avoid extensive parameterizations, and because SAFT models are often used for the prediction of binary or multicomponent systems, for which experimental data is not readily available for one or more of the compounds, pseudo-group contribution methods, in which the pure-component parameters are obtained from mixing rules that combine group-specific parameters, have also been proposed. Those have been applied for some of the most successful SAFT versions such as SAFT-0,[12] SAFT-VR,[12] PC-SAFT,[13–15] and sPC-SAFT,[16,17] and used for the successful description of a wide variety of fluids.[12–20]

Alternatively, instead of applying a GC treatment directly to the pure-component parameters, a more physically meaningful approach would be to develop the underlying theory explicitly in terms of the different functional groups making up the molecule. In this way, the monomeric segments on the chainlike fluid are no longer identical, as occurred with the more common homonuclear SAFT variants. This means that the energy and size characterizing the different monomers are no longer equal to an averaged compound-specific value but can now account for the differences between the different groups present in the fluid. This has led to the development of heteronuclear SAFT models, the most successful being SAFT-γ,[21,22] GC-SAFT-VR,[23] and SAFT-γ-Mie[24] that have been used to the successful description of systems containing different types of compounds including alkanes,[21,23–25] perfluoroalkanes,[26] alkenes,[21,25] alcohols,[21,27] aromatics,[21,23,25,28] ketones,[22,23] acids,[22,25] amines,[22] esters,[23,24,29] and polymers[30].

However, in most of these works with either 'pseudo-group' GC methods or heteronuclear SAFT models, the study of a different family/type of compounds often results in the introduction of a new functional group, a new version of an existing group (e.g. due to changes in its position within the molecule), or a simple refitting of a given functional group. This not only leads to extensive parameter tables that can contain multiple versions of the different groups, but also hinders the assessment of the 'true' transferability of group-specific parameters across different types of compounds.

Hence, the question persists, does the good agreement with experimental data for a multifunctional compound results from a better representation of the effect of the individual groups on the thermodynamic behavior of the fluid? Or the fitting of a new group at each stage of development of a molecular model, even if a small number of adjustable parameters is present, masks any deficiencies on the functional groups that are common to other families?

As an example, when a carboxylic acid is modelled using a heteronuclear SAFT model, $CH_3$, $CH_2$, and COOH groups are involved. The $CH_3$ and $CH_2$ groups are transferred from those previously fitted to the thermodynamic behavior of *n*-alkanes but are they accurately describing the effect of these groups on the carboxylic acids behavior, or are the deficiencies being masked by the parameterization of the COOH group? If the latter is true, then the COOH parameters will be affected by the particular data used in its fitting and will probably fail when used to describe multifunctional molecules that contain COOH groups along with functional groups other than $CH_3$ and $CH_2$.

This work aims at investigating whether the additional complexity of a heteronuclear treatment of the SAFT theory results in an enhanced transferability of the model parameters across different families of compounds, effectively avoiding the need of extensive parameterizations and improving the overall predictive ability of SAFT-type models. For this analysis, one of the most prominent heteronuclear SAFT-type EoSs, SAFT-γ-Mie[24] is used to carry a study on the performance of the EoS to describe the thermodynamic behavior of different families of compounds, a suitable option being the study of three different types of molecules containing a small number of different, but common, functional groups. Due to their importance for the petrochemical industry, the selected compounds are glycols, glymes (or glycol ethers), and the non-ionic alkyl polyglycolether (C*i*E*j*) surfactants. These compounds can be decomposed in four different functional groups (e.g. $CH_3$, $CH_2$, OH, and the ethylene oxide group - $CH_2OCH_2$), only one of which is not found in the well-known *n*-alkanes and alkan-1-ols series. In these compounds, as they are made up of the same functional groups, no additional parameterizations should be carried between them. Moreover, despite the similarity in their chemical structures, these compounds exhibit different properties and a rich variety of phase behaviors when mixed with water, allowing to create a stringent test to the evaluation of the transferability of the parameters on this model.

## 2. Theory: SAFT-γ-Mie EoS

The well-known limitations of the classical cubic EoSs, namely their inability to explicitly account for Lewis acid – Lewis base interactions (e.g. hydrogen bonding) has led to the development of advanced, non-cubic, molecular-based EoSs incorporating statistical thermodynamics concepts, the most successful and popular being those derived from the SAFT theory, proposed by Chapman et al. in the late 80s.[5,6]

Built upon Wertheim's first order thermodynamic perturbation theory (TPT1) proposed a few years earlier,[31–34] SAFT describes molecules as homo-segmented chainlike fluids composed of equally sized spherical monomer beads that exhibit both attractive and repulsive interactions. These segments are tangentially bonded forming chains that may, or may not, associate at specific bonding sites, if the molecular entity being modelled is an associating compound (cf. **Figure 2**). The residual Helmholtz energy of the system, $A^{res}$ is then obtained as a sum of different contributions (or perturbation terms), each accounting for a specific structural or energetic effect as illustrated in **eq. 1**: a reference or monomer term, $A^{ref}$ that accounts for both the repulsive and attractive interactions between the monomeric segments; the chain formation term, $A^{cha}$ , and an association term, $A^{assoc}$, that explicitly account for short-range and highly directional forces such as hydrogen-bonding being acknowledged at the most important feature of such advanced models when compared with the classical methods.

$$\frac{A^{res}}{Nk_BT} = \frac{A}{Nk_BT} - \frac{A^{ideal}}{Nk_BT} = \frac{A^{ref}}{Nk_BT} + \frac{A^{cha}}{Nk_BT} + \frac{A^{assoc}}{Nk_BT} + \cdots \quad (1)$$

Furthermore, given the additive nature of SAFT, additional terms can be added, depending on the systems nature and complexity, the most common example being the inclusion of a polar term to account for additional electrostatic effects. While both the chain and association terms have their foundation on Wertheim's TPT1 theory, different reference terms have been chosen to account for the physical interactions between monomers, especially in the attractive contribution, leading to the development of a number of different SAFT variants including the original SAFT,[6] CK-SAFT,[35] soft-SAFT,[36] SAFT-VR,[37] and PC-SAFT, in both its original[38] and simplified version[39], or even the special case of CPA that replaces the reference and chain terms by a cubic EoS such as Peng-Robinson or Soave-Redlich-Kwong.[40]

As previously mentioned, most of these EoSs apply a homonuclear approach to treat molecules as homo-segmented chains. Since all the beads constituting the molecule are equal, i.e. have the same energy and size, each compound has its own individual set of molecular parameters, representing "averaged" mean values that are able to capture most physical features of the compound being studied, although not explicitly accounting for the heterogeneity of the molecules, or the different functional groups present in their chemical structure.

Aiming, at the development of more physically realistic models, of enhanced predictive ability and transferability, heteronuclear treatments of the SAFT-theory, coupled with GC methods, have been proposed in the framework of SAFT-type EoSs. As illustrated in **Figure 1**, where homonuclear and heteronuclear coarse-grained (CG) models for propanoic acid are schematically represented, these type of EoSs allow for the study of chainlike molecules built with beads having different characteristics that can then be associated to key functional groups making up the molecules. The clear advantage of these approaches is that, as in most GC-based approaches, once parameters have been determined for a wide range of functional groups (whose parameterization can be carried using a broader range of experimental data, and not only the pure fluid of interest), one can easily predict the thermodynamic behavior of a virtually unlimited number of compounds without further fitting.

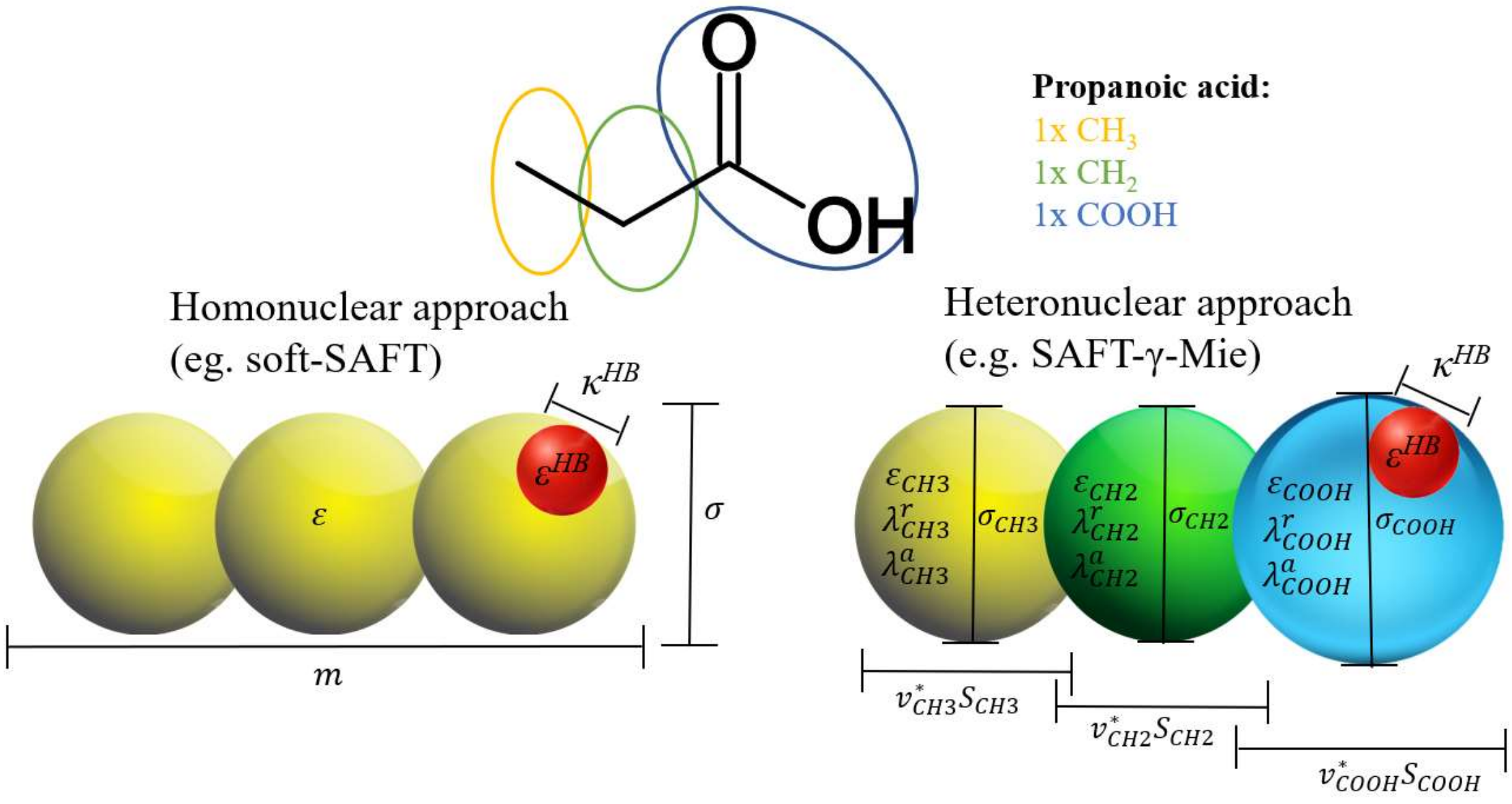

**Figure 1.** CG models of propanoic acid under an homonuclear and heteronuclear approach.

One of the most prominent heteronuclear SAFT-type EoSs is the SAFT-γ-Mie approach[24], where molecules are represented as associating heteronuclear chains of fused spherical segments. Within the SAFT-γ-Mie framework, each functional group $k$ is represented as a number of spherical segments, $v_k^*$, (note that in the example provided in **Figure 1**, all groups are composed of only one segment, i.e. $v_k^* = 1$), and every pair of segments in the mixture ($k$, $l$) are assumed to interact according to a Mie potential of variable range, according to **eq. 2**, where $r_{kl}$ is the distance between the centers of the two segments, $\sigma_{kl}$ is the segment diameter, $\varepsilon_{kl}$ is the depth of the potential well, and $\lambda_{kl}^r$ and $\lambda_{kl}^a$ are the repulsive and attractive exponents of the segment-segment potential.

$$\Phi_{k,l}^{Mie} = \frac{\lambda_{kl}^r}{\lambda_{kl}^r - \lambda_{kl}^a}\left(\frac{\lambda_{kl}^r}{\lambda_{kl}^a}\right)^{\frac{\lambda_{kl}^a}{\lambda_{kl}^r-\lambda_{kl}^a}} \varepsilon_{kl}\left[\left(\frac{\sigma_{kl}}{r_{kl}}\right)^{\lambda_{kl}^r} - \left(\frac{\sigma_{kl}}{r_{kl}}\right)^{\lambda_{kl}^a}\right] \tag{2}$$

Considering that the model contemplates the use of fused segments, instead of tangentially bonded ones, an additional parameter, known as the shape factor, $S_k$, is required to reflect the fraction of the segment that effectively contributes to the system's free energy. Furthermore, as in the more classical homonuclear approaches, associating forces are modelled by incorporating a number of short-range square-swell association sites, which are, in this case, placed in any segment of the functional group exhibiting such interactions. Hence, for each 'associating group' the number and type of association sites must be specified, and any interaction allowed to occur between an associating site type $a$ in group $k$ and a site type $b$ in group $l$ has to be characterized by two additional parameters: the association energy, $\varepsilon_{kl,ab}^{HB}$, and the bonding volume, $K_{kl,ab}^{HB}$.

In summary, any non-associating functional group $k$ becomes fully defined once the values for $v_k^*$, $S_k$, $\sigma_{kk}$, $\varepsilon_{kk}$, $\lambda_{kk}^r$ and $\lambda_{kk}^a$ are known. If the compound is associating, the number of association sites of each type has to be defined along with the energy, $\varepsilon_{kk,ab}^{HB}$, and volume, $\kappa_{kl,ab}^{HB}$, characterizing each type of site-site interaction. From these parameters, $v_k^*$ can usually be defined a priori based on the size/structure of the functional group being considered, and $\lambda_{kk}^a$ is usually kept constant to its default value of 6, unless for very specific cases, being usually excluded from the parameterization procedure.

Within a heteronuclear approach such as SAFT-γ-Mie, the unlike interaction parameters are required not only to treat mixtures, as occurs in classical SAFT models, but also for pure component calculations, unless the compound of interest is made up of only one functional group, as it is the case for water or methanol. For the unlike interactions, the values of $\sigma_{kl}$, $\varepsilon_{kl}$, $\lambda^r_{kl}$, $\lambda^a_{kk}$, $\varepsilon^{HB}_{kl,ab}$, and $\kappa^{HB}_{kl,ab}$ have to be specified. These parameters can be obtained from combining rules employing the group-specific parameters, namely:

$$\sigma_{kl} = \frac{\sigma_{kk} + \sigma_{ll}}{2} \quad (3)$$

$$\varepsilon_{kl} = \frac{\sqrt{\sigma^3_{kk}\sigma^3_{ll}}}{\sigma^3_{kl}}\sqrt{\varepsilon_{kk}\varepsilon_{ll}} \quad (4)$$

$$\lambda_{kl} = 3 + \sqrt{(\lambda_{kk} - 3)(\lambda_{ll} - 3)} \quad (5)$$

$$\varepsilon^{HB}_{kl,ab} = \sqrt{\varepsilon^{HB}_{kk,aa}\varepsilon^{HB}_{ll,bb}} \quad (6)$$

$$\mathrm{K}^{HB}_{kl,ab} = \left(\frac{\sqrt[3]{\mathrm{K}^{HB}_{kk,aa}} + \sqrt[3]{\mathrm{K}^{HB}_{ll,bb}}}{2}\right)^3 \quad (7)$$

These combining rules are known to provide a good initial estimate of the necessary unlike parameters; nonetheless, for a higher accuracy of the model when used to describe experimental systems, available experimental data is typically used to refine the values obtained using **eqs. 3-7**, especially those characterizing the unlike dispersive and associative energies ($\varepsilon_{kl}$ and $\varepsilon^{HB}_{kl,ab}$). Such parameters can be obtained from fitting to the experimental data along the group-specific parameters and, contrarily to homonuclear models, the unlike interaction parameters can be obtained from pure-component data alone, assuming the selected compounds contain all the functional groups under study.

The detailed expressions for each of the terms present in **eq. 1**, required for the calculation of the residual Helmholtz energy of the system are not presented here for sake of simplicity, but can be found in the original publication of the SAFT-γ-Mie EoS.[24] Once its value its obtained, a wide range of thermodynamic properties, including pressure, and the chemical potential, required for phase equilibrium calculations, can be easily obtained, using appropriate derivatives and ideal gas integrals.

## 3. Results and discussion

### 3.1. *n*-alkanes

Due to the importance of $CH_3$ and $CH_2$ groups for the modelling of a wide variety of compounds, including those of interest to this work, the parameterization proposed by Papaioannou et al.[24] in the original SAFT-γ-Mie paper is here revisited. Both the group-specific parameters and the unlike interactions between the methyl and methylene functional groups, reported in **Table 1** and **Table 2**, were obtained by fitting to experimental data for vapor pressure, saturated liquid density, and single-phase density of $n$-alkanes from ethane to $n$-decane, and exhibit percentage average absolute relative deviations (%AARD) of 1.55 and 0.59 for vapor pressure and saturated liquid density data, respectively.

As shown in **Figures S1-S2** in Supporting Information, the performance of the model in describing the vapor pressure and the saturated liquid density is indeed remarkable except for ethane and $n$-propane. This is however expected since they are the shortest oligomers of the $n$-alkanes series and the thermodynamic behavior of the first members of a homologous series of compounds can deviate from their higher homologous. In the same article, Papaioannou et al.[24] showed that, using these parameters, good predictions for the thermodynamic behavior of $n$-alkanes of longer chain lengths (up to C30) could be obtained, including a good description of second-order derivative properties such as the isobaric and isochoric heat capacities ($c_p$ and $c_v$), speed of sound ($u$), isothermal compressibility ($k_T$) and isobaric expansivity ($\alpha_P$). Additionally, as it is shown in **Figure S3**, a good prediction of vaporization enthalpies can also be obtained for $n$-alkanes up to at least 20 carbon atoms.

Another important test for the validity of the $CH_3$ and $CH_2$ parameters is the modelling of binary mixtures of $n$-alkanes. Papaioannou et al.[24] showed that, using these parameters, excellent descriptions of both the VLE and excess properties, namely the excess speed of sound and excess molar volumes, could be obtained. Particularly remarkable, is the good agreement with the experimental data obtained for a very asymmetrical system ($n$-propane + $n$-hexacontane). In addition, it is shown in **Figure S4**, in Supporting Information, that the model is able not only to successfully describe both isothermal and isobaric vapor-liquid equilibrium (VLE) of binary $n$-alkane mixtures, but also their densities, even at high pressures.

For all these reasons, the parameters previously estimated for the methyl and methylene groups (and their unlike interaction) by Papaioannou et al. [24] will be used throughout this work.

**Table 1.** Group specific parameters used in the SAFT-γ-Mie EoS calculations.

| Group | $v_k$ | $S_k$ | $\lambda_{kk}^{r}$ | $\lambda_{kk}^{a}$ | $\sigma_{kk}$[Å] | $\varepsilon_{kk}/k_B[K]$ | $NS_{k,H}$ | $NS_{k,e1}$ | $NS_{k,e2}$ | *Reference* |
|---|---|---|---|---|---|---|---|---|---|---|
| **$CH_3$** | 1 | 0.57255 | 15.050 | 6 | 4.0772 | 256.77 | - | - | - | [24] |
| **$CH_2$** | 1 | 0.22932 | 19.871 | 6 | 4.8801 | 473.39 | - | - | - | [24] |
| **$CH_2OH$** | 2 | 0.58538 | 22.699 | 6 | 3.4054 | 407.22 | 1 | 2 | - | [41] |
| **$H_2O$** | 1 | 1.00000 | 17.020 | 6 | 3.0063 | 266.68 | 2 | 2 | - | [25] |
| **OH** | 1 | 0.82154 | 10.316 | 6 | 2.8965 | 350.00 | 1 | 2 | - | This work |
| **EOa** | 2 | 0.55560 | 11.860 | 6 | 3.8050 | 277.39 | 0 | 0 | 1 | This work |
| **EOb** | 1 | 0.55386 | 10.000 | 6 | 4.8330 | 367.68 | 0 | 0 | 1 | This work |
| **EOg** | 2 | 0.55560 | 20.011 | 6 | 3.805 | 394.26 | 0 | 0 | 1 | This work |
| **$CH_3Oa$** | 1 | 0.6860 | 10.05 | 6 | 3.9910 | 473.50 | 0 | 0 | 1 | This work |
| **$CH_3Og$** | 1 | 0.9650 | 10.26 | 6 | 3.5690 | 493.48 | 0 | 0 | 1 | This work |

**Table 2.** Unlike physical interaction parameters used in the SAFT-γ-Mie EoS calculations.

| Group *k* | Group *l* | $\varepsilon_{kl}/k_B[K]$ | $\lambda_{kl}^{r}$ | *Reference* |
|---|---|---|---|---|
| **$CH_3$** | **$CH_2$** | 350.77 | CR | [24] |
| **$CH_3$** | **$CH_2OH$** | 333.20 | CR | [41] |
| **$CH_3$** | **$H_2O$** | 358.18 | 100.00 | [41] |
| **$CH_3$** | **OH** | 249.96 | CR | This work |
| **$CH_2$** | **$H_2O$** | 423.63 | 100.00 | [41] |
| **$CH_2$** | **$CH_2OH$** | 423.17 | CR | [41] |
| **$CH_2$** | **OH** | 352.53 | CR | This work |
| **$CH_2OH$** | **$H_2O$** | 353.37 | CR | [41] |
| **OH** | **$H_2O$** | 323.55 | CR | This work |
| **EOa** | **CH3** | 383.60 | CR | This work |
| **EOa** | **CH2** | 296.00 | CR | This work |
| **EOa** | **OH** | 472.44 | CR | This work |
| **EOa** | **$H_2O$** | 289.44 | CR | This work |
| **EOb** | **CH3** | 443.01 | CR | This work |
| **EOb** | **CH2** | 262.27 | CH2 | This work |
| **EOb** | **OH** | 351.21 | CR | This work |
| **EOb** | **$H_2O$** | 261.54 | CR | This work |
| **EOg** | **CH3** | 269.10 | CR | This work |
| **EOg** | **CH2** | 325.24 | CR | This work |
| **EOg** | **OH** | 451.94 | CR | This work |
| **EOg** | **$H_2O$** | 340.65 | CR | This work |
| **CH3Oa** | **CH3** | 458.70 | CR | This work |
| **CH3Oa** | **EO** | 232.41 | CR | This work |
| **CH3Oa** | **$H_2O$** | 279.30 | CR | This work |
| **CH3Og** | **CH3** | 478.68 | CR | This work |
| **CH3Og** | **EO** | 293.53 | CR | This work |
| **CH3Og** | **$H_2O$** | 335.74 | CR | This work |

### 3.2. Water and *n*-alkane + water mixtures

As mentioned in the introduction, the ethylene oxide-containing compounds that are the cornerstone of this work present a rich and varied phase behavior when in aqueous solutions. Therefore, the modelling of their mixtures with water will also be considered throughout this work to evaluate the performance of the SAFT-γ-Mie EoS and the transferability of parameters, across different families of compounds. Consequently, it is important to have an accurate and robust parameterization of water and of their unlike interactions with the groups of interest.

Dufal et al.[25] were the first to propose a parameterization of the water molecule, in the framework of SAFT-γ-Mie, along with parameters for the unlike interactions between water and a considerable number of functional groups. Of special relevance to this work are the water group specific parameters (reported in **Table 1** and **Table 3**), and the parameters characterizing their unlike interaction with the methyl and methylene groups, reported in **Table 2**. Although details concerning the experimental data used in the fitting of such parameters were not provided, the model was shown to provide a good description of the very low mutual solubilities exhibited in water + benzene and water + $n$-hexane systems, and a good prediction of the excess isobaric heat capacity ($C_p^E$) of acetone + water mixtures.

In a subsequent study by the same group,[41] the authors found that, using this parameterization, a good description of the $n$-alkane solubility in the water-rich phase could only be obtained for medium chain length hydrocarbons. Therefore, the parameters characterizing the unlike interaction between the alkyl groups and water were refined, considering both the $\varepsilon_{kl}$ and $\lambda_{kl}^r$ as adjustable parameters. While keeping the water molecule parameters equal to those previously reported,[25] three-phase vapor-liquid-liquid equilibrium (VLLE) solubility data for aqueous mixtures of $n$-pentane and $n$-octane, over an extended temperature range, and the coexisting liquid compositions at 298 K, along the three-phase region, for mixtures of $n$-hexane and $n$-decane mixtures were used to obtain the unlike parameters mediating the $H_2O$-$CH_3$ and $H_2O$-$CH_2$ interactions, reported in **Table 2**.

As shown in **Figure 2**, these new parameters allowed for a considerable improvement on the description of the $n$-alkanes solubility in water, while preserving a similar accuracy when describing the water solubility in $n$-alkanes (cf. Hutacharoen et al.[41]).

**Table 3.** Site-site hydrogen-bonding parameters.

| **Group *k*** | **Group *l*** | **Site *a* of group *k*** | **Site *b* of group *l*** | $\varepsilon_{kl,ab}^{HB}[K]$ | $\kappa_{kl,ab}^{HB}/[\text{Å}^3]$ | ***Reference*** |
|---|---|---|---|---|---|---|
| **$H_2O$** | **$H_2O$** | H | $e_1$ | 1985.4 | 101.69 | [25] |
| **$CH_2OH$** | **$CH_2OH$** | H | $e_1$ | 2097.9 | 62.309 | [41] |
| **$CH_2OH$** | **$H_2O$** | H | $e_1$ | 621.68 | 425.00 | [41] |
| **$CH_2OH$** | **$H_2O$** | $e_1$ | H | 2153.2 | 147.40 | [41] |
| **OH** | **OH** | H | $e_1$ | 2491.9 | 21.340 | This work |
| **OH** | **$H_2O$** | H | $e_1$ | 1690.5 | 37.640 | This work |
| **OH** | **$H_2O$** | $e_1$ | H | 2064.83 | 179.98 | This work |
| **EOa** | **OH** | $e_2$ | H | 2523.3 | 21.340 | This work |
| **EOa** | **$H_2O$** | $e_2$ | H | 1381.95 | 315.32 | This work |
| **EOb** | **OH** | $e_2$ | H | 745.29 | 678.20 | This work |
| **EOb** | **$H_2O$** | $e_2$ | H | 2461.63 | 15.01 | This work |
| **EOg** | **OH** | $e_2$ | H | 1196.0 | 21.340 | This work |
| **EOg** | **$H_2O$** | $e_2$ | H | 1319.83 | 61.18 | This work |
| **$CH_3Oa$** | **$H_2O$** | $e_2$ | H | 2793.22 | 14.96 | This work |
| **$CH_3Og$** | **$H_2O$** | $e_2$ | H | 3161.61 | 15.00 | This work |

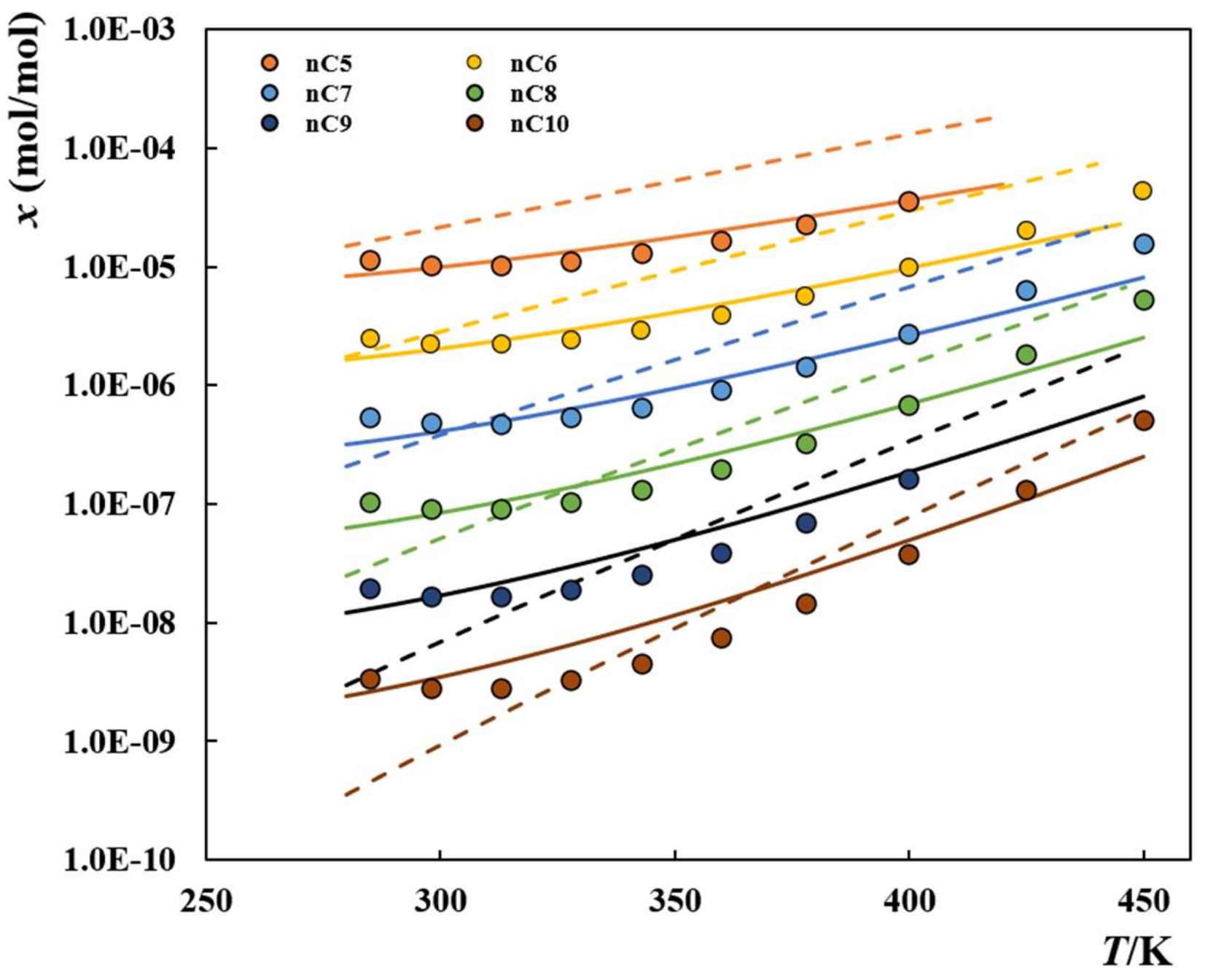

**Figure 2.** Solubility of n-alkanes in water. Symbols represent experimental data[42–47] while the solid lines and dashed lines represent the SAFT—Mie results using the unlike interaction parameters with alkyl groups proposed by Hutacharoen et al.[41] and Dufal et al.[25], respectively.

### 3.3. Alkan-1-ols

Once the alkyl groups are established within the GC framework, they can be transferred to the modelling of more complex compounds, through the successive introduction of additional groups. The compounds of interest to this work contain one or two terminal alcohol groups, also present in the much simpler alkan-1-ols. Evidently, alcohol groups can be modelled either as the hydroxyl group itself (i.e. OH), or along with the methylene group to which it is bonded, considering it as a $CH_2OH$ group.

To the best of our knowledge, only Hutacharoen et al.[41] have proposed a parameterization for the study of 1-alcohols in the framework of SAFT-γ-Mie. In their work, a new $CH_2OH$ group, whose parameters are reported in **Tables 1-3**, was introduced. As reported in **Table 1**, the $CH_2OH$ group proposed contains two identical segments and, being an associating group, a 3B association scheme, according to nomenclature proposed by Huang and Radosz[48], was assigned. The new group contains two association site types: two sites type $e_1$ representing the two lone electron pairs on the oxygen atom, and one site type $H$, representing the hydrogen atom of the hydroxyl group, with only unlike site-site interactions being allowed. The parameters obtained for the $CH_2OH$ and its unlike interactions with $CH_3$ and $CH_2$ groups were obtained by fitting to the vapor pressure and saturated-liquid density of pure $n$-alkan-1-ols from ethanol to $n$-decan-1-ol along with the LLE data for $n$-tetradecane + ethanol.

Although this model is able to successfully describe both the pure-component properties of alkan-1-ols and the phase equilibria of binary mixtures $n$-alkanes + alkan-1-ols and water + alkan-1-ols, a new parameterization of the hydroxyl group, considering it as an OH group, instead of a $CH_2OH$, is proposed in this work. The idea is not to replace the parameterization proposed by Hutacharoen and coworkers,[41] but rather to be able to test the two alternatives in the modelling of other families of compounds, and evaluate the transferability of the hydroxyl group across the different compounds of interest.

To parameterize the new OH group, only one segment was considered (i.e. $\upsilon_k^* = 1$) and, similarly to the $CH_2OH$ group available in the literature, the 3B association scheme was assigned to represent its hydrogen bonding character. The attractive exponent, $\lambda_{kk}^a$, was fixed to the default value of 6 as commonly done for most groups,[25] while the remaining unknown parameters,$S_k$, $\lambda_{kk}^r$, $\sigma_{kk}$, $\varepsilon_{kk}/k_B$, $\varepsilon_{kl,ab}^{HB}$, $\kappa_{kl,ab}^{HB}$ , and the parameters characterizing the unlike dispersive energy with the $CH_3$ and $CH_2$ groups were regressed from the vapor pressures and saturation liquid densities of pentan-1-ol, hexan-1-ol, octan-1-ol, and decan-1-ol, along with the VLE of butan-1-ol + $n$-decane mixtures. The OH-group specific parameters obtained are reported in **Table 1** and **3**, while those mediating the unlike interactions with alkyl groups are given in **Table 2**, respectively. Comparing to the $CH_2OH$ parameters reported in the literature, the new OH parameters show a much lower repulsive exponent of the Mie potential, which is coherent with the lower OH-OH, OH-$CH_2$ and OH-$CH_3$ dispersive energies obtained.

This difference may be due to the fact that, contrarily to the parameterization carried out by Hutacharoen and coworkers,[41] pure fluid data for ethanol and propan-1-ol were not included in the experimental dataset used for the fitting, which could force the repulsive exponent of the Mie potential towards higher values. Instead, it was opted to exclude them from the dataset as the shortest oligomers of a given homologous series can have a slightly different behavior than their higher chain length homologues, since the global effect of the presence of a terminal alcohol group can be affected by the very small alkyl chain attached to it. Consequently, as reported in **Table 4**, the deviations between the model calculations and the VLE experimental data for pure ethanol are higher than those obtained using the $CH_2OH$ group available in literature[41]. On the other hand, for most of the remaining oligomers, including propan-1-ol, the new parameterization yields a better description of the data.

To enhance the robustness of the OH group parameterization, the VLE of butan-1-ol and + *n*-decane mixture was also considered in the fitting procedure. Then, as all the necessary parameters were available, the model was used to predict the VLE of 25 different binary alkan-1-ols + *n*-alkanes mixtures. The deviations from the experimental data are reported in **Table 5**, along with those obtained if a $CH_2OH$ group is considered and, as can be gauged from the table, both models have a similar accuracy, although the new OH group shows lower deviations for 17 out of the 25 mixtures investigated. Nevertheless, in most cases the differences between the two models are negligible, and the highest deviations are usually observed on the dew pressures of systems where experimental data is only available on a very narrow region of the phase diagram (e.g. butan-1-ol + *n*-dodecane and tetradecan-1-ol + *n*-undecane).

**Table 4.** Deviations between the SAFT-γ-Mie EoS results and the experimental data[49] for the vapor pressure, saturation liquid density and vaporization enthalpy of pure alkan-1-ols, using the alcohol group parameterized as an OH (this work) or a $CH_2OH$[41].

| Compound | *T-range (K)* | $\%AARD_{OH}$ $p^*/\rho_L/H_{vap}$ | $\%AARD_{CH2OH}$ $p^*/\rho_L/H_{vap}$ |
|---|---|---|---|
| **ethanol** | 200-510 | 5.06/2.60/2.25 | 2.27/2.64/1.59 |
| **propan-1-ol** | 200-530 | 5.85/1.48/2.01 | 6.91/1.35/1.58 |
| **butan-1-ol** | 220-550 | 3.39/1.26/5.07 | 7.51/1.44/5.14 |
| **pentan-1-ol** | 240-580 | 4.95/1.61/6.67 | 3.37/2.03/6.01 |
| **hexan-1-ol** | 250-600 | 2.90/2.38/5.79 | 3.20/2.93/5.31 |
| **octan-1-ol** | 280-640 | 2.12/2.12/10.92 | 6.00/2.90/10.24 |
| **decan-1-ol** | 290-660 | 2.56/2.24/8.61 | 3.62/2.93/8.55 |

**Table 5.** Binary alkan-1-ols + *n*-alkanes systems studied in this work, using the SAFT-γ-Mie EoS and the correspondent deviations from the experimental data modelling the hydroxyl groups as an OH (this work) or a $CH_2OH$ group.[41]

| System | $T$ (K) | $P$ (kPa) | $\%AARD_{OH}$ | $\%AARD_{CH2OH}$ | *Exp. Ref.* |
|---|---|---|---|---|---|
| **ethanol + propan-1-ol** | 303.15 | 12-16.02 | 7.58 | 3.86 | [50] |
| **ethanol + $nC_4$** | 323-423 | 28-3800 | 2.69/4.30 | 4.00/7.23 | [51] |
| **ethanol + $nC_6$** | 328-473 | 46-3484 | 2.23/2.75 | 5.72/5.54 | [52] |
| **ethanol + $nC_7$** | 303.15 | 4-11 | 1.48 | 3.32 | [50] |
| **ethanol + $nC_{11}$** | 333-353 | 0.5-106 | 7.15 | 8.68 | [53] |
| **propan-1-ol + $nC_6$** | 338-348 | 21-138 | 2.53/4.64 | 6.08/6.37 | [54] |
| **propan-1-ol + $nC_7$** | 303.15 | 5-11 | 1.05 | 4.09 | [50] |
| **propan-1-ol + $nC_{11}$** | 333-353 | 0.5-51 | 7.95 | 10.5 | [53] |
| **butan-1-ol + $nC_5$** | 333-393 | 8-918 | 3.37/6.21 | 2.27/7.65 | [55] |
| **butan-1-ol + $nC_6$** | 323 | 21-55 | 5.56/1.85 | 1.03/0.48 | [56] |
| **butan-1-ol + $nC_7$** | 313.15 | 2.5-13.2 | 2.43/2.13 | 1.96/1.71 | [57] |
| **butan-1-ol + $nC_8$*** | 382-399 | 101.3 | 0.838* | 1.79* | [58] |
| **butan-1-ol + $nC_8$** | 313-373 | 2.4-76 | 4.48/2.49 | 6.06/4.42 | [59,60] |
| **butan-1-ol + $nC_9$** | 323 | 4.7-5.7 | 3.15/2.96 | 4.48/3.25 | [56] |
| **butan-1-ol + $nC_{10}$** | 358-388 | 5-92.5 | 3.56/2.95 | 5.45/4.37 | [61] |
| **butan-1-ol + $nC_{11}$** | 353-373 | 1.7-52.6 | 5.51 | 5.96 | [53] |
| **butan-1-ol + $nC_{12}$** | 313.15 | 0.05-2.5 | 5.67/19.1 | 3.93/16.8 | [57] |
| **pentan-1-ol + $nC_5$** | 303 | 0.4-82 | 5.28 | 3.55 | [62] |
| **pentan-1-ol + $nC_7$** | 348-368 | 7-92 | 0.86/0.77 | 3.14/1.08 | [63] |
| **pentan-1-ol + $nC_9$*** | 406-424 | 101.3 | 0.381* | 1.23* | [64] |
| **hexan-1-ol + $nC_6$*** | 342-422 | 101.3 | 1.233* | 1.152* | [65] |
| **hexan-1-ol + $nC_7$*** | 371-428 | 101.3 | 1.873* | 2.981* | [65] |
| **octan-1-ol + $nC_6$** | 313.15 | 8.8-35.9 | 3.55 | 2.53 | [56] |
| **dodecan-1-ol + $nC_6$** | 298.15 | 7-19.7 | 1.96 | 1.57 | [56] |
| **dodecan-1-ol + $nC_{11}$** | 393-413 | 0.5-20.5 | 3.61 | 3.65 | [53] |
| **tetradecan-1-ol + $nC_{11}$** | 393-413 | 0.1-20.5 | 2.64/14.4 | 3.60/14.1 | [53] |

*Isothermal VLE – the deviations provided are an average absolute deviation in $T$ (K)
When deviations are reported as a pair of values they are related to bubble/dew pressures.

### 3.4. Alkan-1-ols + water mixtures

With the OH parameters and their unlike interactions with the alkyl groups defined, the model can be used to investigate the phase equilibria of alkan-1-ols + water mixtures. Contrarily to the *n*-alkanes + water mixtures where the presence of weak attractive interactions dominates the systems behavior, in the case of alkan-1-ols + water mixtures, hydrogen bonding also plays an important role with the fluid phase behavior of these mixtures resulting from a delicate balance between the two types of interactions.

Consequently, while all *n*-alkanes are fully immiscible in water (mutual solubilities are very low), methanol, ethanol and propan-1-ol are fully miscible in water, and thus the phase diagram of their aqueous mixtures, consists of a simple VLE. On the other hand, from butan-1-ol onwards, alkan-1-ols are only partly miscible in water (the water solubility in the alcohol-rich phase being much higher than the solubility of the alcohol in water-rich phase), and thus the isobaric temperature-composition ($T$-$x$) phase diagrams usually consist of a vapor-liquid-liquid equilibria (VLLE). To be able to model these mixtures, the unlike interaction parameters (dispersive energy and hydrogen bonding parameters) between the alcohol groups and water must be first estimated from experimental data. As previously done by Hutacharoen et al.[41] for the $CH_2OH$ group, the unknown parameters between the new OH group and water were regressed from the LLE data of the octan-1-ol + water mixture and the final values are reported in **Table 2** and **Table 3**. Afterwards, the parameters were used to predict the VLE and/or LLE data for alkan-1-ols + water mixtures from ethanol to dodecan-1-ol, an example being the satisfactory prediction of the VLLE phase diagram of the pentan-1-ol + water mixture shown in **Figure 3**.

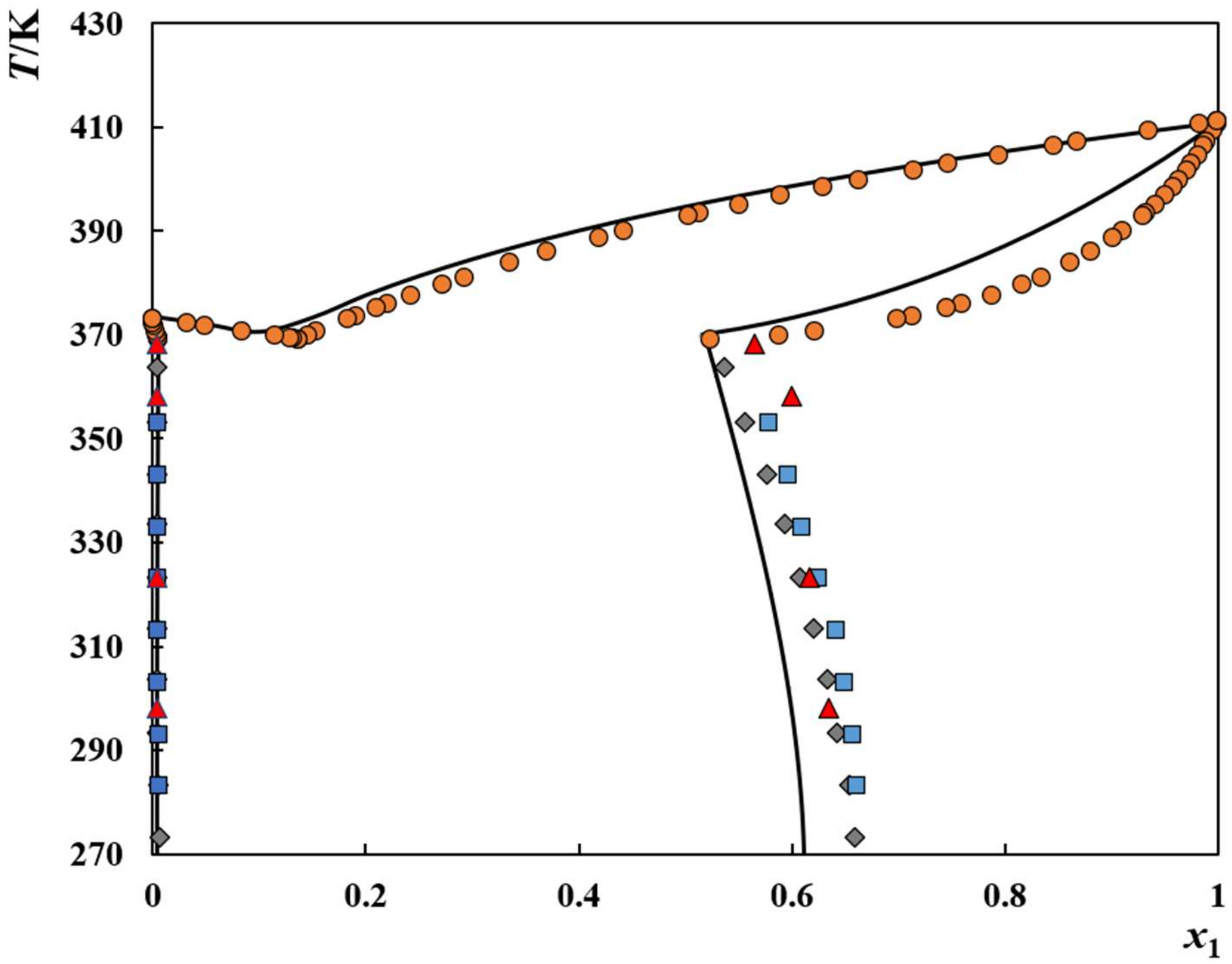

**Figure 3.** Vapor-liquid-liquid equilibria of pentan-1-ol (1) + water (2) at atmospheric pressure. Symbols represent experimental data[66–69] while the solid lines depict the SAFT-γ-Mie EoS results using the new OH group.

The performance of the new parameterization for the alcohol group, when used to describe alkan-1-ol + water mixtures, was further compared with the CG model that considers a $CH_2OH$ group, and the deviations from the experimental data are reported in **Table 6**. As can be seen in **Figure 4**, both approaches have a similar performance when used to describe the mutual solubilities in such mixtures. The deviations from the LLE experimental data show that the new approach performs better in the alcohol phase and in the water-rich phase for the shorter alcohols, while the $CH_2OH$ approach performs better in the water-rich phase of longer alcohols. In terms of VLE, both models have a similar accuracy, and reasonable results can be obtained, especially if considering that VLE data was not used in the parameterization of the unlike parameters between the alcohol and water groups.

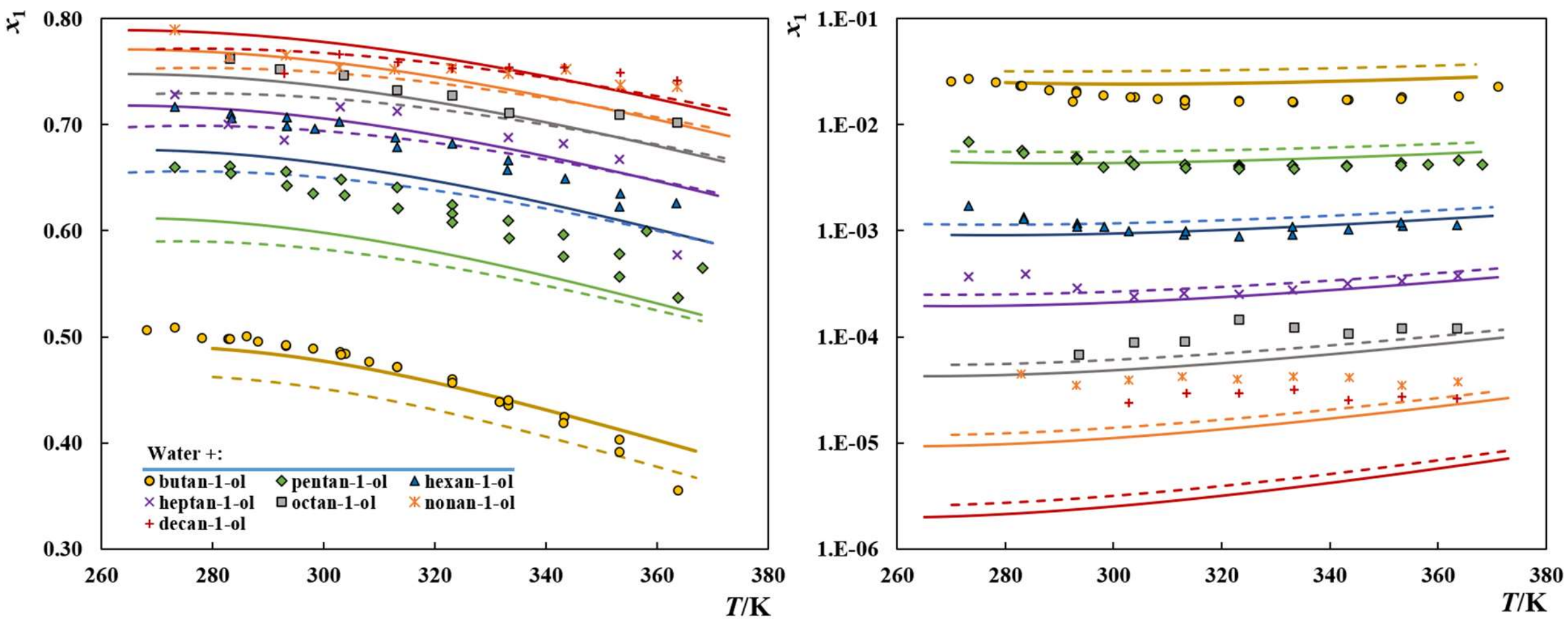

**Figure 4.** Liquid-liquid equilibria of alkan-1-ol (1) + water mixtures. (left) alcohol rich phase (right) aqueous phase. Symbols represent experimental data while the solid and dashed lines represent the SAFT-γ-Mie results using an OH or a $CH_2OH$ group, respectively.

**Table 6.** Binary alkan-1-ols + water systems studied in this work, using the SAFT-γ-Mie EoS and the correspondent deviations from the experimental data modelling the hydroxyl groups as an OH (this work) or a $CH_2OH$ group.[41]

| | ***Vapor-Liquid Equilibrium*** | | | |
|---|---|---|---|---|
| **System** | ***P* (kPa)** | $AAD(T)/K_{OH}$ | $AAD(T)/K_{CH2OH}$ | ***Exp. Ref.*** |
| **ethanol + water** | 13-33 | 3.01/2.89 | 3.12/3.15 | [70] |
| **propan-1-ol + water** | 30-100 | 5.21/4.80 | 5.43/5.00 | [71] |
| **butan-1-ol + water** | 101.3 | 3.67/3.16 | 3.94/3.36 | [72] |
| **pentan-1-ol + water** | 101.3 | 4.99/1.87 | 5.06/1.96 | [66] |
| **hexan-1-ol + water** | 101.3 | 7.44/1.87 | 7.14/1.86 | [73] |
| | ***Liquid-Liquid Equilibrium*** | | | |
| | ***T-range* (K)** | $\%AARD(z)_{OH}$ | $\%AARD(z)_{CH2OH}$ | ***Exp. Ref.*** |
| **butan-1-ol + water** | 273-363 | 3.09/34.5 | 6.16/77.0 | [74–76] |
| **pentan-1-ol + water** | 273-363 | 5.03/19.5 | 7.20/38.2 | [66–69] |
| **hexan-1-ol + water** | 273-363 | 5.76/21.0 | 6.46/31.3 | [67,77] |
| **heptan-1-ol + water** | 273-363 | 3.13/18.0 | 3.43/16.7 | [67] |
| **octan-1-ol + water** | 283-363 | 1.27/42.1 | 2.52/25.5 | [67,78,79] |
| **nonan-1-ol + water** | 273-363 | 1.58/63.9 | 2.40/53.0 | [76] |
| **decan-1-ol + water** | 293-363 | 1.42/86.4 | 1.08/82.1 | [76] |

In VLE calculations the pair of deviations presented are related to the Bubble/Dew temperatures and are reported as an average deviation in terms of temperature. In the LLE calculations, the values reported are for the alcohol/aqueous phase and are reported as %ARD.

### 3.5. α, ω - alkanediols

As mentioned above the main goal of this work is to evaluate the transferability of parameters and the predictive ability of heteronuclear SAFT-type EoSs. This means to evaluate the performance of group-specific and unlike interaction parameters across different families, whose study do not require the simultaneous introduction and fitting of new additional groups. Having parameterized the alkyl groups ($CH_3$ and $CH_2$), and the alcohol group that, as seen in the previous sections, can be modelled either using an OH or a $CH_2OH$ bead, one class of compounds that can be studied without further parameterizations is the α,ω – alkanediols.

Experimental data is available for the pure fluid densities and vapor pressures for 1,2-ethanediol, 1,3-propanediol, 1,4-butanediol, and 1,5-pentanediol.[49] Hence, the parameters previously reported in **Tables 1-3**, were used in a predictive manner to calculate these properties. Such calculations are illustrated in **Figure 4** for 1,3-propanediol and 1,5-pentanediol, while the deviations from the experimental data for all the four compounds are given in **Table 7**. As can be observed in **Figure 4**, both approaches fail to accurately describe these fundamental properties of pure α,ω – alkanediols. Although using the new OH group a reasonable description of density data could be obtained, especially at low temperatures, the description of the vapor pressures is very poor for all the compounds, using both approaches, with deviations from the experimental data ranging between 17.5 and 48.3%.

These results are a first indication of the limited transferability of group and unlike parameters across different families, suggesting that the effect of the hydroxyl groups in a compound molecular structure, or at least, the additive effect of having multiple units of these groups in a given compound may not be correctly accounted for by using multiple alcohol groups in an heteronuclear EoS.

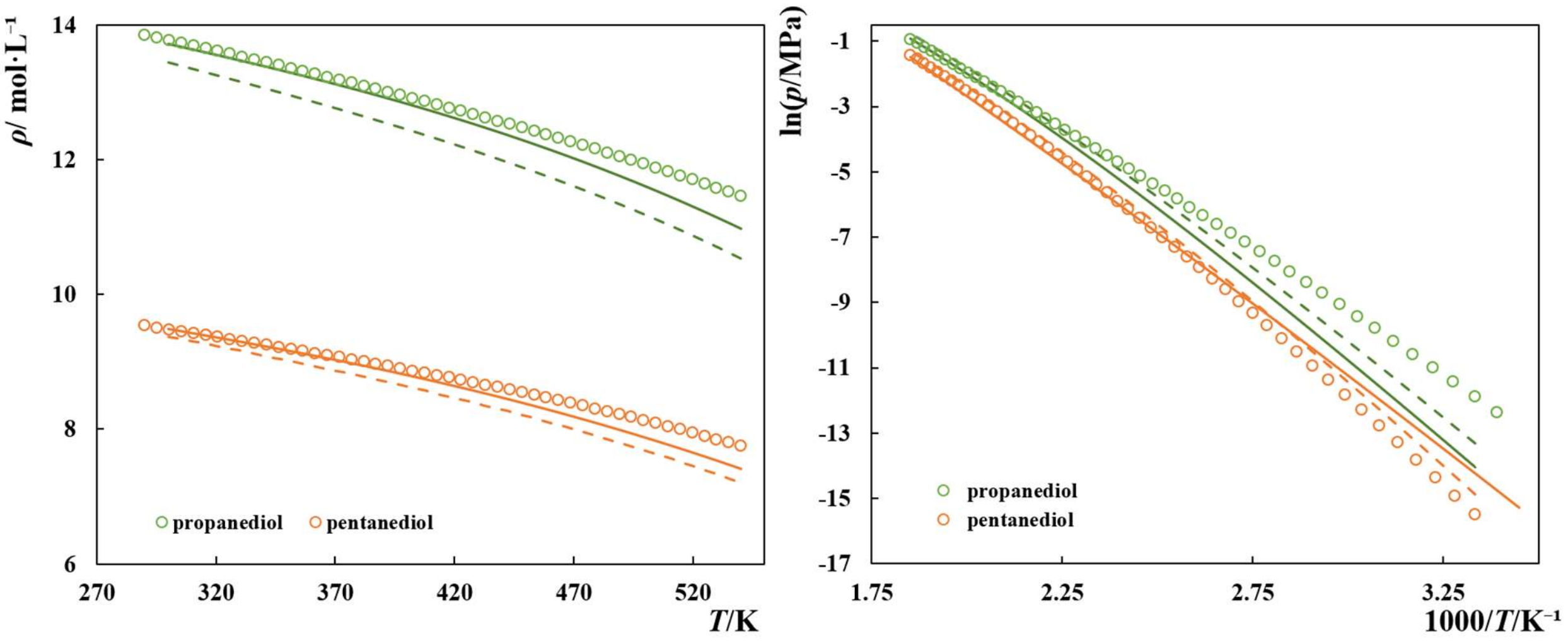

**Figure 5.** Densities and vapor pressures for α, ω-alkanediols. Symbols represent experimental data[49] while the solid and dashed lines represent the SAFT-γ-Mie EoS results using an OH or a $CH_2OH$ to describe the alcohol groups, respectively.

**Table 7.** Deviations between SAFT-γ-Mie results and the experimental data for pure α,ω - alkanediols.

| Compound | *T-range (K)* | $\%AARD_{OH}$ $p^*/\rho_L$ | $\%AARD_{CH2OH}$ $p^*/\rho_L$ |
|---|---|---|---|
| **1, 2 – ethanediol** | 280-640 | 48.3/0.80 | 33.9/5.45 |
| **1, 3 – propanediol** | 300-540 | 43.1/1.46 | 31.8/4.46 |
| **1, 4 – butanediol** | 300-540 | 17.5/0.93 | 20.2/3.37 |
| **1, 5 - pentanediol** | 300-540 | 26.8/1.55 | 31.5/3.49 |

### 3.6. Pure glycols and their mixtures with *n*-heptane

The alkyl and alcohol groups having been defined, the performance of SAFT-γ-Mie EoS, when used to describe the thermodynamic behavior of compounds containing ethylene oxide (EO) groups, can be investigated. Among such compounds, glycols are widely used in the oil and gas industry for several purposes such as gas hydrate inhibitors or as dehydration agents in natural gas streams.[80] Although their chemical structure is typically expressed as H-$(OCH_2CH_2)_nOH$, three different approaches were considered in this work to subdivide the glycol molecules into groups. These approaches are schematically represented in **Figure 6** for triethylene glycol: approach A is based on the chemical formula presented above so that each glycol consists on a terminal hydroxyl group, and $n$ EO groups, each described as an -$OCH_2CH_2$ unit. On the other hand, approach B and C attempt to retain the identity of both end groups by considering that a glycol has two terminal alcohol groups and, consequently two $CH_2$ groups and $n-1$ EO groups represented as either a $CH_2OCH_2$ (approach B) or -$CH_2CH_2O$ (approach C). Given the nature of the SAFT-γ-Mie EoS, approach B and C are indistinguishable, since the connectivity between the different groups is not explicitly considered in the theory, as conveniently explained in the work of Papaioannou et al.[24] On the other hand, from approach A results that ethylene glycol (MEG) is no longer seen as an α,ω – alkanediol but is rather divided into an EO group and an additional OH group.

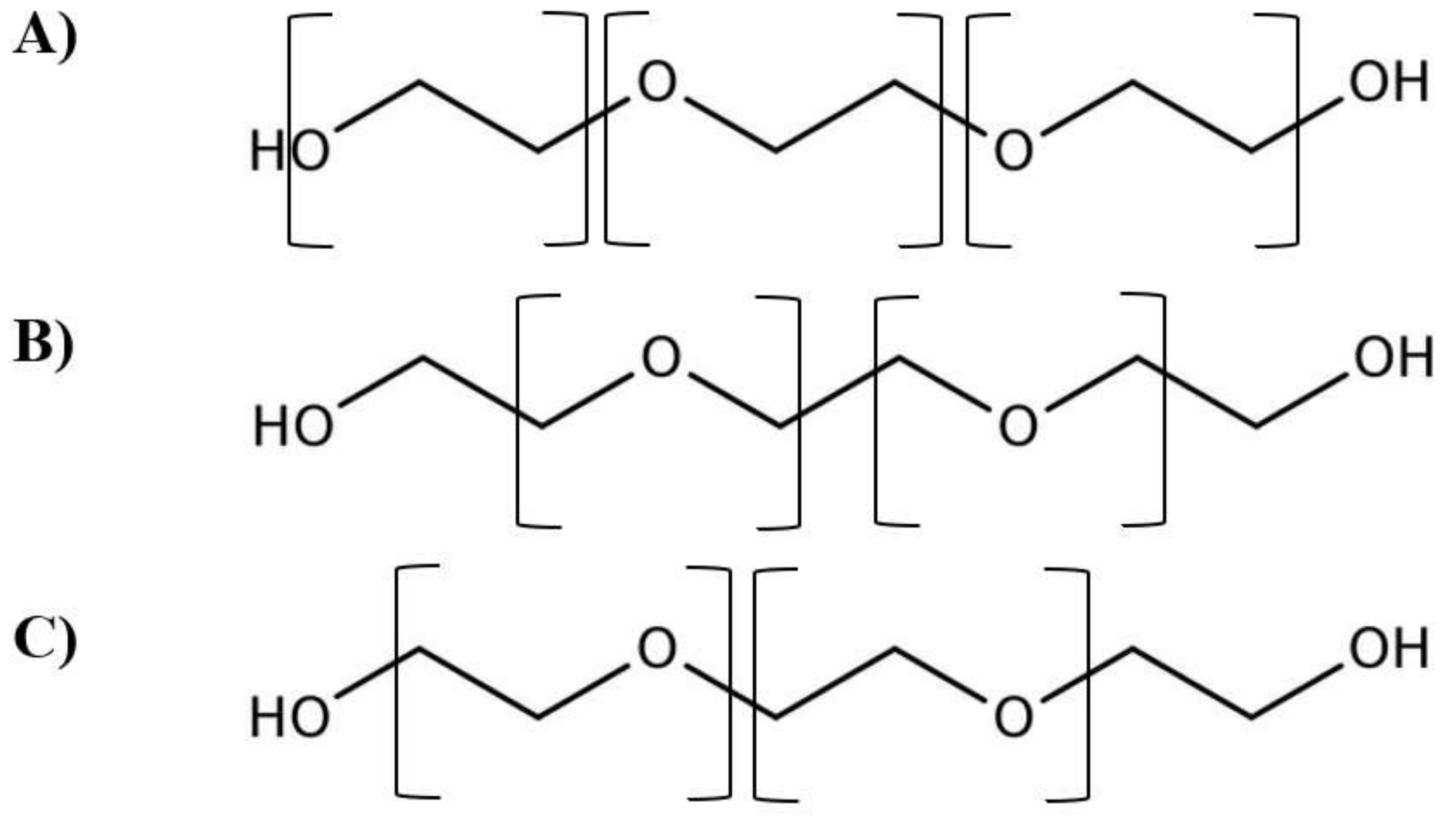

**Figure 6.** Approaches considered to subdivide the glycol molecules in the framework of SAFT-γ-Mie groups.

Considering the slightly better performance of the OH group proposed in this work, compared with the literature $CH_2OH$ group to describe alcohol groups, shown in previous sections, and because only the OH group is coherent with approach A, the OH group will be adopted as the primary option throughout the rest of this work to represent alcohol groups. Nonetheless, assessments of the performance of the model using a $CH_2OH$ group were carried out but no significant advantages were found for the systems investigated.

To the best of our knowledge, this is the first time that the EO group parameterization is carried out using the SAFT-γ-Mie EoS. Considering $\lambda^{a}_{EO-} = 6$, as previously suggested for most groups, the remain unknown parameters necessary to define an EO group are $\upsilon^{*}_{EO}$, $S_{EO}, \lambda^{r}_{EO-EO}, \sigma_{EO-}$ , $\varepsilon_{EO-}$ . Furthermore, to model a pure glycol oligomer the unlike interaction parameters with the remaining groups are also required, namely the unlike dispersive energies between the EO groups and the alkyl or alcohol groups, i.e. $\varepsilon_{EO-CH}$ , $\varepsilon_{EO-CH2}$, and $\varepsilon_{EO-OH}$. Moreover, the EO group, due to the lone electron pairs on the oxygen atom is able to establish hydrogen bonds so their associative behavior can be modelled by considering the existence of one or two 'negative' association sites type '$e_2$' that can interact for example with the 'H' site on the alcohol groups. Hence, two additional parameters are required, namely the energy and volume of association between said sites, $\varepsilon^{HB}_{EO-\ \ ,e2-H}$, $\kappa^{HB}_{EO-OH,e2-H}$.

All the unknown parameters were obtained by fitting to experimental data available in literature for pure glycol oligomers and their mixtures with *n*-alkanes, namely the vapor pressures and saturated liquid densities of pure diethylene glycol (DEG), triethylene glycol (TriEG), the saturated liquid densities of tetraethylene glycol (TeEG) and the LLE data for the systems DEG + *n*-heptane and TriEG + *n*-heptane. The vapor pressures of the TeEG were not considered in the parameterization dataset due to very high uncertainty of the data available in the DIPPR database, partially related with the very low volatility of TeEG at low to medium temperatures, as previously discussed by other authors.[80]

Due to the very high number of unknown parameters involved, a sequential parameterization procedure was considered. In an initial step the association parameters between EO and OH groups were set equal to those mediating the hydrogen bonding between OH groups. Simultaneously, only the pure-component data was considered so that the unlike interaction parameter with the $CH_3$ group was not necessary in this first step. The initial results suggested that for approach A, EO groups should be modelled using two segments ($\upsilon^{*}_{k} = 2$), since all the tests carried using only one segment resulted in shape factors always tending to the upper limit of 1. On the contrary, for approach B, the results suggested the use of a single, but larger, bead to model the EO group, i.e. with a higher value of sigma. Once the number of segments was defined, further optimization iterations were carried in order to significantly narrow the range of optimal values for $S_{EO}, \lambda^{r}_{EO-E}$ , $\sigma_{EO-}$ in each approach.

In a subsequent stage, the selected mixture data was included in the parameterization dataset and all the unknown parameters were fitted using the optimal parameter ranges obtained in the previous step, considering zero, one or two association sites of type '$e_2$' per EO group. Concerning the number of association sites, the results obtained suggest that the associating behavior of EO cannot be neglected but using one or two association sites result in a very similar performance of the model. Therefore, considering that such a GC-based transferable model should be suitable for a wide range of chain length/molecular weight, allowing for example the study of polyethylene glycols that can have a considerable number of EO groups, only one association site, type '$e_2$' per group was considered in order to decrease the mathematical burden of calculating the association term in future polymer calculations.

The final values of the parameters obtained for the EO group following the two approaches, 'EOa' and 'EOb' are reported in **Tables 1-3**, while the results of the fitting are shown in **Figure 7** for the pure fluid data (including the prediction of TeEG vapor pressures and predictions for MEG) and in **Figure 8** for the LLE of glycols + *n*-heptane systems.

As can be seen in **Figure 7**, and by the deviations from the experimental data reported in **Table 8**, both approaches show significant deviations from the experimental vapor pressures, especially at lower temperatures were the pressure values become very small. For the saturation liquid densities, the deviations are much smaller but only approach B is able to describe density data with an accuracy similar to that obtained using homonuclear SAFT models.[9]

The two modelling approaches were further used to predict the *pρT* data previously reported for glycols in the 283-363 K and 0.1-95 MPa temperature and pressure ranges,[9] and the deviations from the experimental data are also presented in **Table 8**. The results show that approach B is, as previously seen, much better than approach A to describe glycols densities, even at high pressures as shown in **Figure 9**, but the deviations exhibited by approach A (see data description in **Figure S6**) have decreased considerably when compared to the saturated liquid densities, suggesting that the pressure effect on density is well captured also by this approach but there are increasing deviations with temperature that are more relevant on the saturated liquid densities due to the broader temperature range considered.

Concerning the LLE of mixtures with *n*-heptane shown in **Figure 8**, approach A shows a remarkable accuracy in describing the TriEG + *n*-heptane system (deviations reported in **Table 9**), while the deviations increase considerably for the mixtures containing DEG or TeEG, especially in the glycol-rich phase. For approach B the deviations are very high in all cases, although a better performance than with approach A was obtained in the glycol-rich phase of the mixtures with DEG and TeEG.

At this stage it is not possible to confidently determine which of the two approaches is better, although the results suggest that approach A might be better only for specific glycol chain lengths, namely for the particular case of the TriEG while approach B might be better overall, especially in the description of pure component properties. Therefore, both modelling approaches will be considered in the next section for the study of glycol + water mixtures.

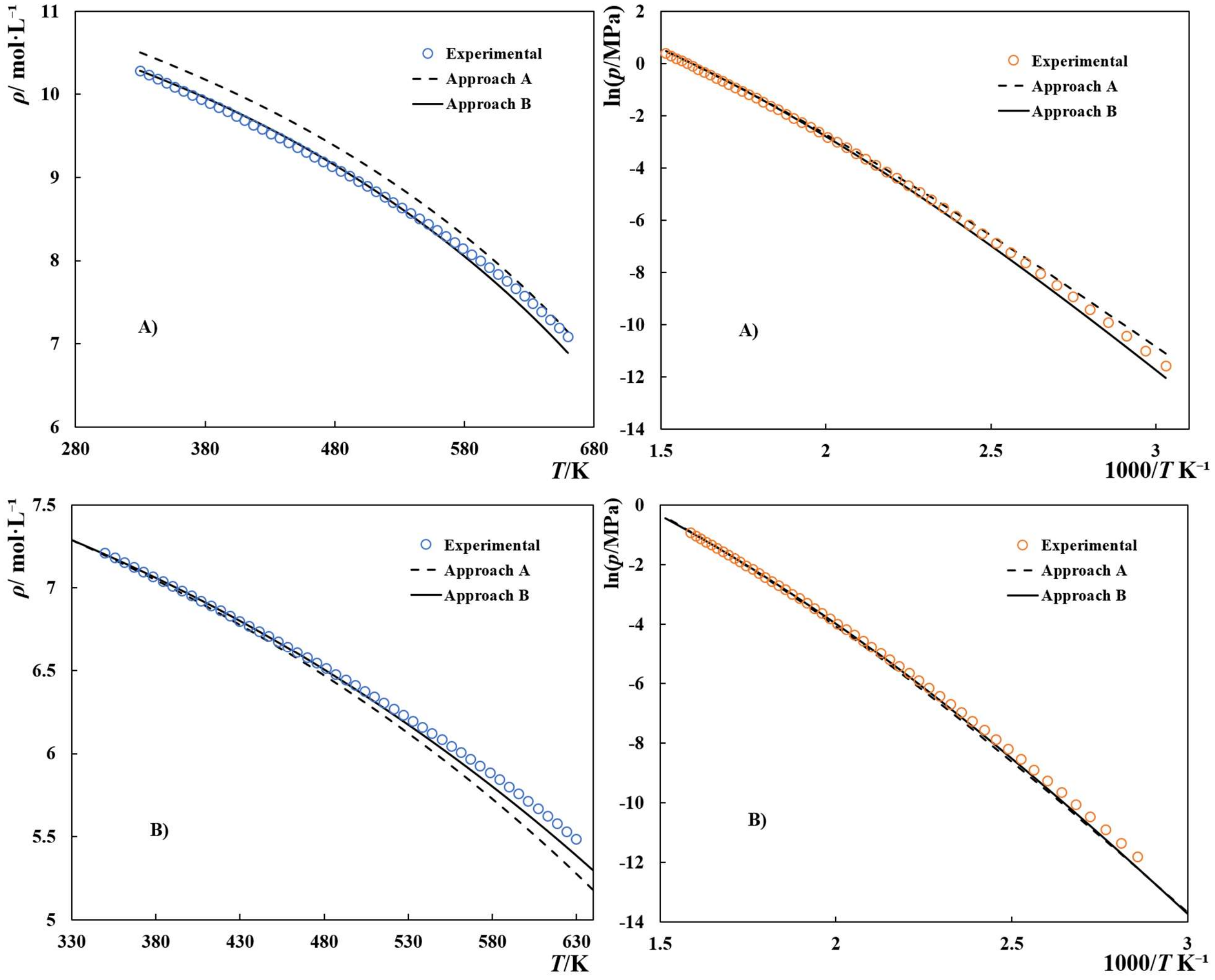

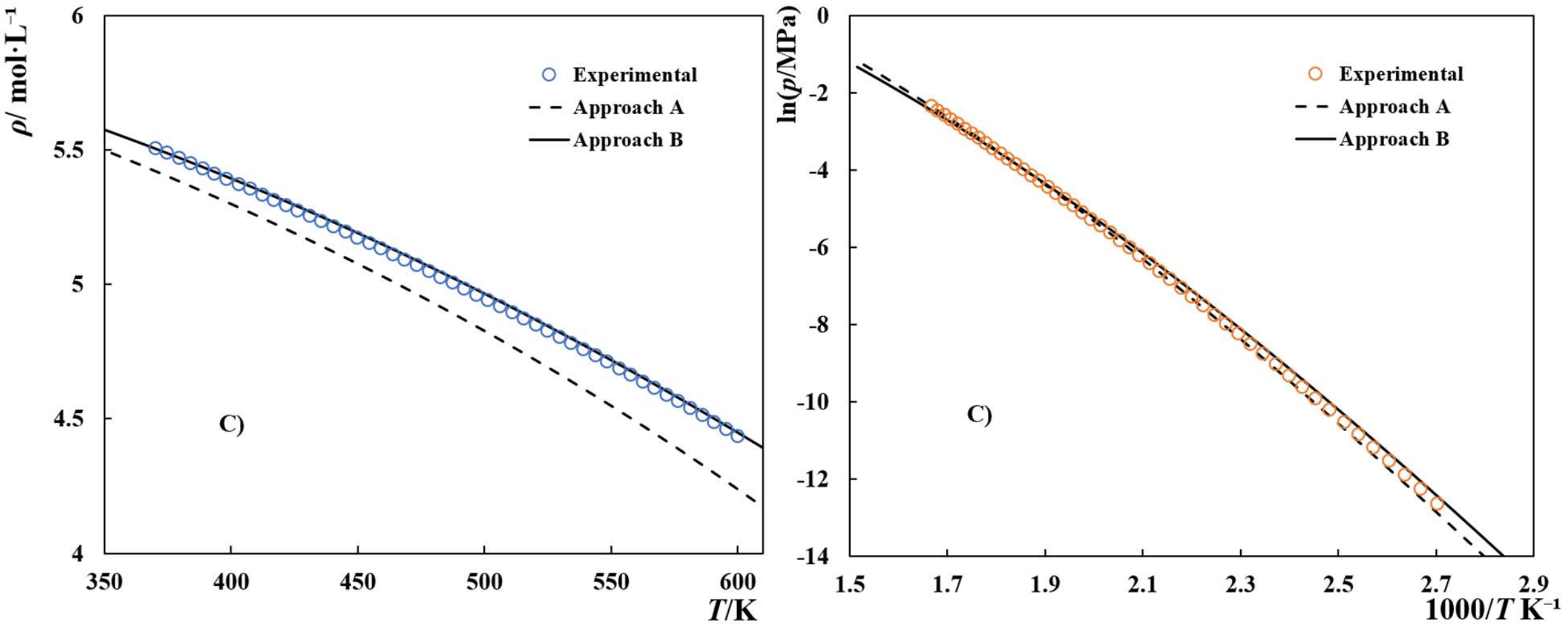

**Figure 7.** Saturation liquid densities and vapor pressures of pure glycol oligomers: A) DEG; B) TriEG; C) TeEG. Symbols represent experimental data[49] while the solid and dashed lines represent the SAFT-γ-Mie results.

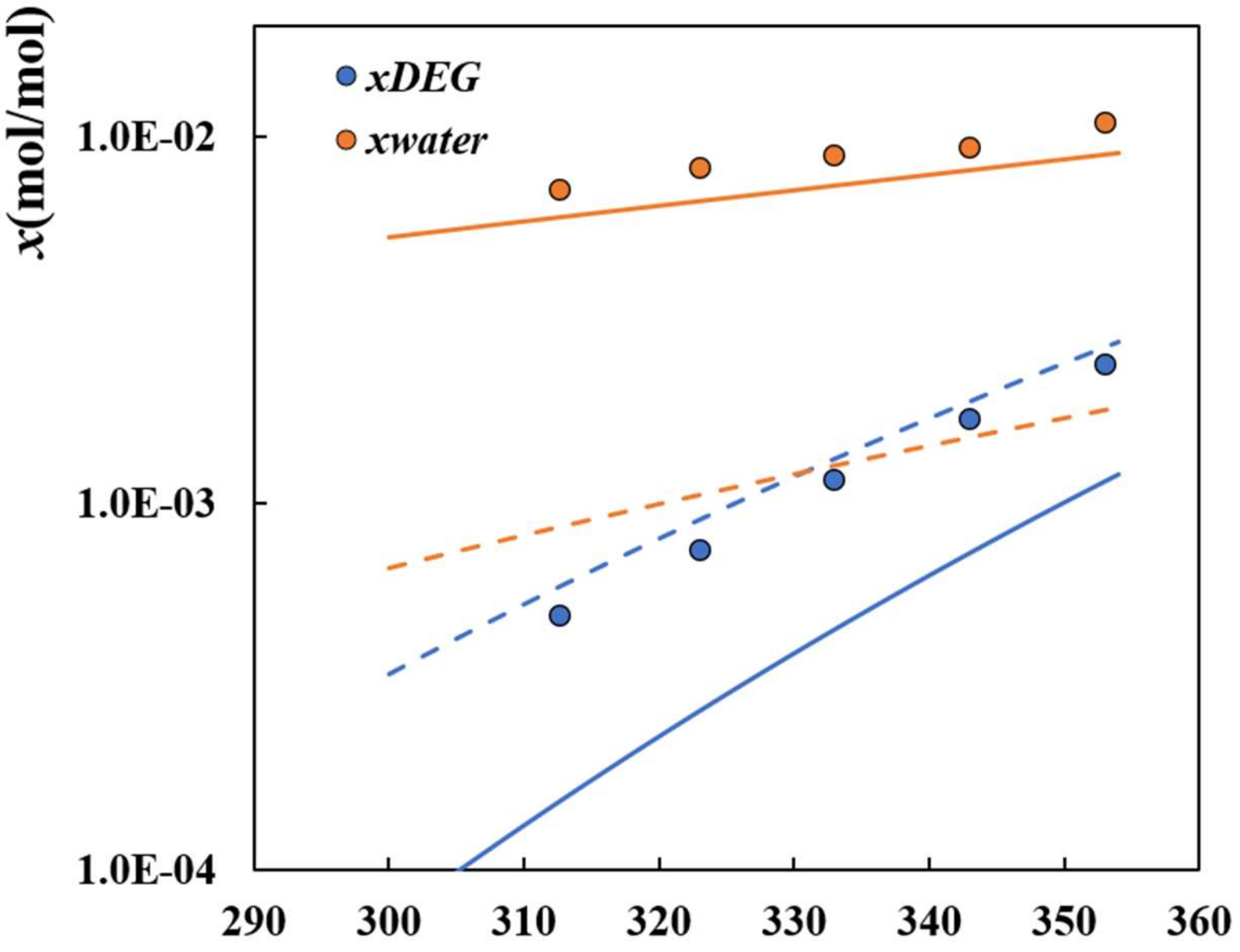

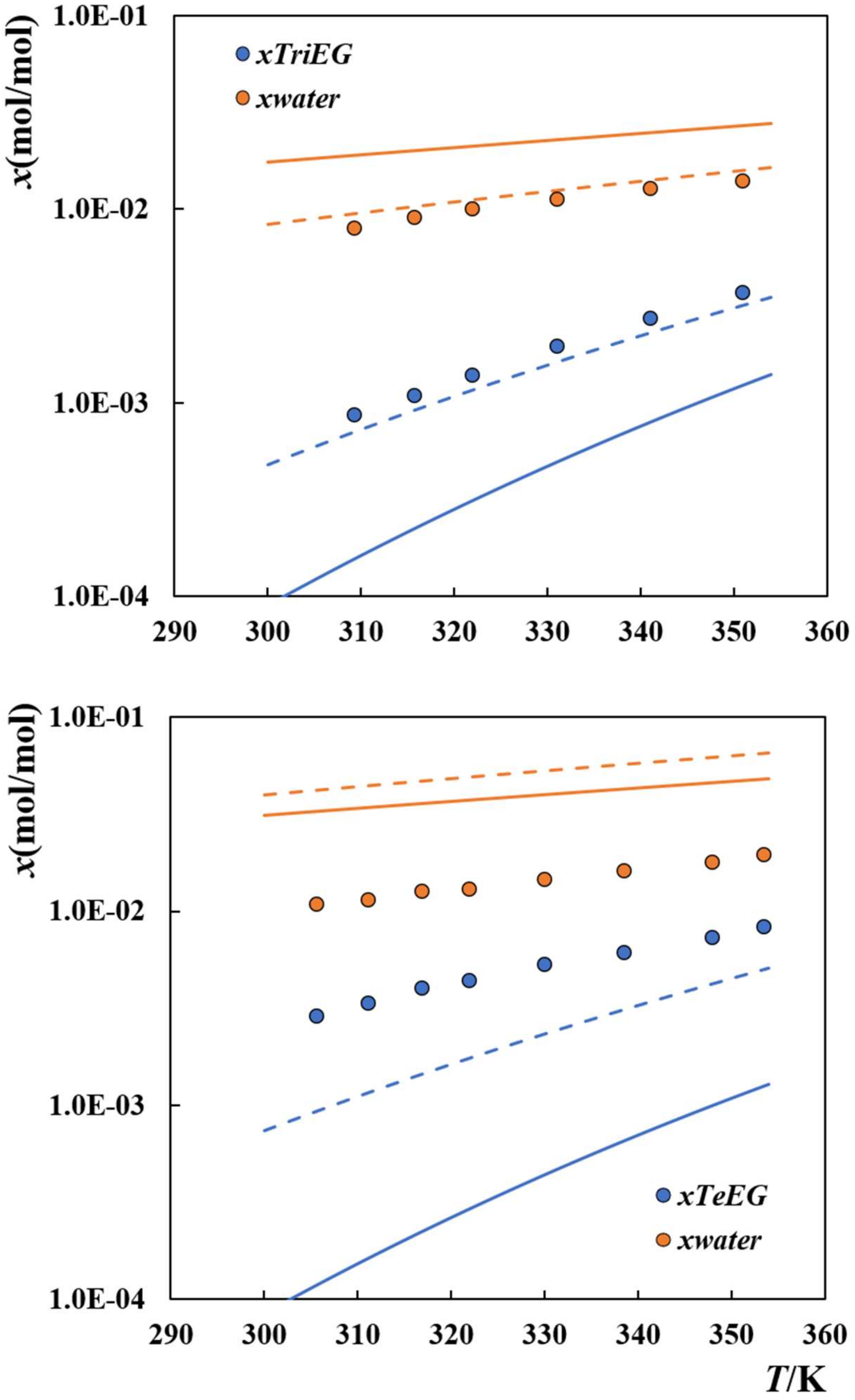

**Figure 8.** Liquid-liquid equilibrium of glycol + n-heptane systems. Symbols represent experimental data[81] while the dashed and solid lines represent the SAFT-γ-Mie results using approach A and approach B, respectively.

**Table 8.** Deviations between the experimental data for pure glycols[49] and the SAFT-γ-Mie results in %AARD.

| | Approach A | | | Approach B | | |
|---|---|---|---|---|---|---|
| | $\rho_L$ | $p^*$ | $p\rho T$ | $\rho_L$ | $p^*$ | $p\rho T$ |
| **MEG** | 7.66 | 58.09 | - | 1.02 | 42.09 | - |
| **DEG** | 2.36 | 14.52 | 2.105 | 0.77 | 13.44 | 0.188 |
| **TriEG** | 1.25 | 14.11 | 0.242 | 0.52 | 9.24 | 0.342 |
| **TeEG** | 2.54 | 5.91 | 0.850 | 0.28 | 11.16 | 0.520 |
| **PeEG** | - | - | 1.620 | - | - | 0.574 |
| **HeEG** | - | - | 2.097 | - | - | 0.634 |

**Table 9.** Deviations between the SAFT-γ-Mie results and the experimental data for the LLE of glycol (1) + n-heptane (2) systems[81] expressed in %AARD. $x_1^{II}$ represents the molar fraction of glycol in the alkane-rich phase, while $x_2^{I}$ represents the water mole fraction in the glycol rich-phase.

| | Approach A | | Approach B | |
|---|---|---|---|---|
| | $x_1^{II}$ | $x_2^{I}$ | $x_1^{II}$ | $x_2^{I}$ |
| **DEG** | 16.34 | 85.50 | 60.07 | 16.27 |
| **TriEG** | 15.74 | 13.74 | 75.19 | 111.44 |
| **TeEG** | 55.20 | 266.8 | 91.34 | 178.20 |

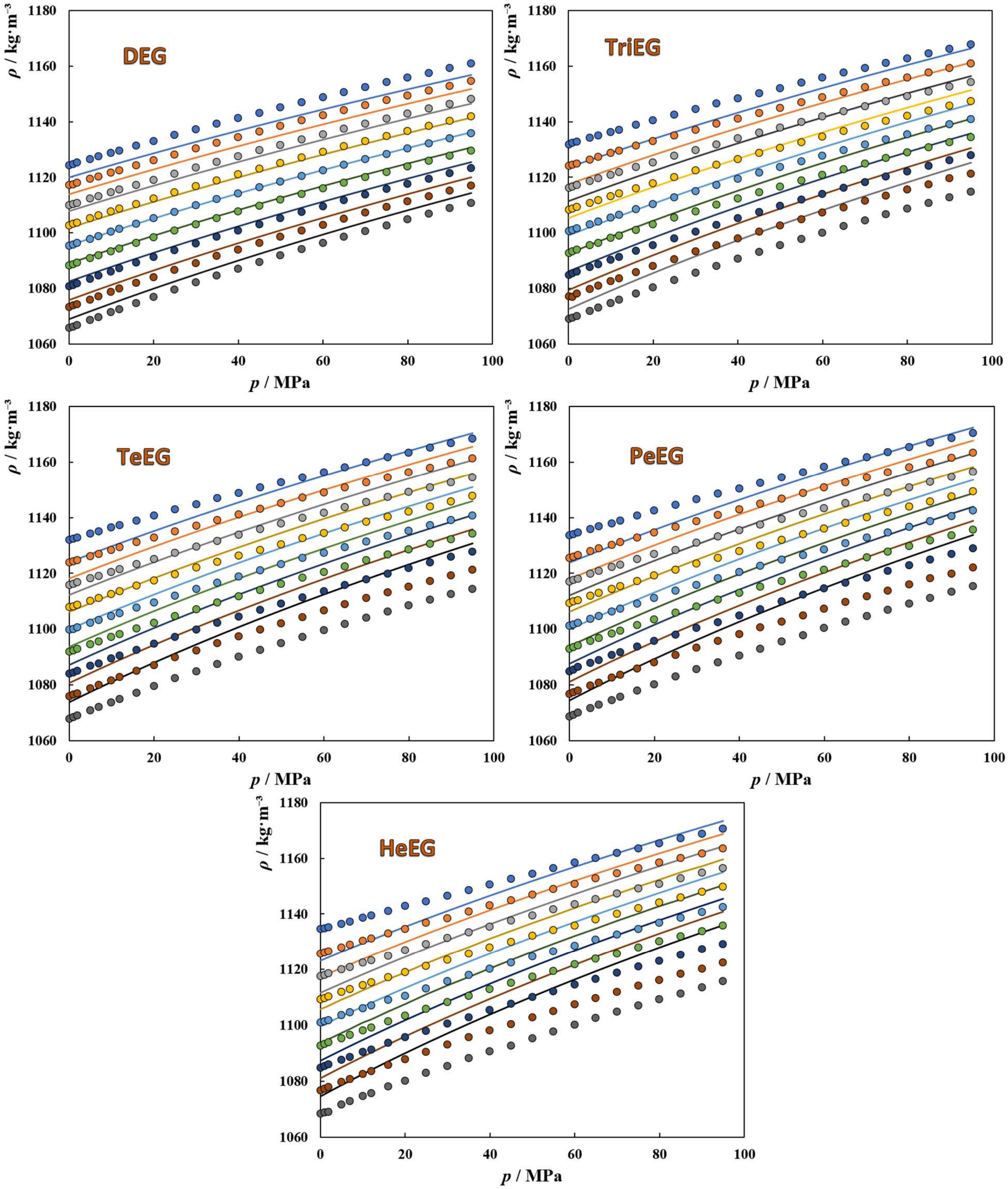

**Figure 9.** High-pressure liquid densities of pure glycols. Symbols represent experimental data[9] while the solid lines depict the SAFT-γ-Mie results following the approach B.

### 3.7. Glycols + water mixtures

Having estimated the parameters for the new EO group and their unlike interactions with the alkyl and hydroxyl groups, it is now possible to investigate the performance of the model when used to describe glycol aqueous solutions, in order to further compare the performance of the two modelling approaches considered in the previous section. For such calculations, the parameters characterizing the unlike interactions between the EO groups and water have to be defined, namely $\varepsilon_{EO-H2O}$, $\varepsilon^{HB}_{EO-H2O,e2-H}$, and $\kappa^{HB}_{EO-H2O,e2-H}$. These parameters were here regressed from available experimental data for the VLE of DEG, TriEG, and TeEG aqueous solutions at 0.1 MPa, and used to predict the behavior at two other pressures, 0.05 and 0.07 MPa.

The results of the fitting procedure are shown in **Figure 10**, and the deviations from the experimental data, reported as average absolute deviation (K), are provided in **Table 10**. As can be observed in the figure, approach A shows a better agreement with the experimental boiling temperatures for the three systems, especially for DEG + water for which, a quantitative agreement using approach B would require a further tuning of the $\varepsilon^{HB}_{EO-H2\ \ ,e2-H}$ parameter to a value different than those used for their higher chain length homologues. Moreover, using approach B, one can observe that while increasing the glycols chain length, the model goes from overestimating to underestimating the boiling temperatures, i.e. from overestimating to underestimating the water-glycol interactions.

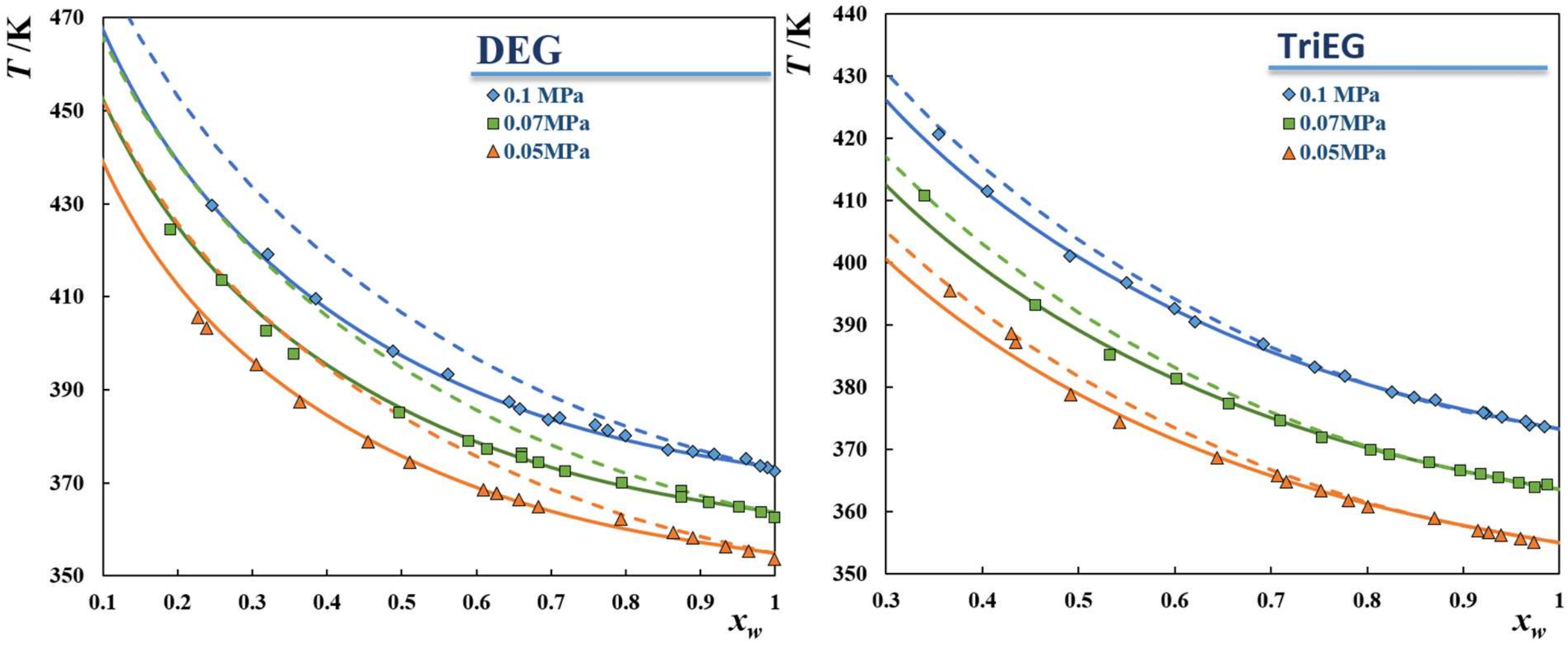

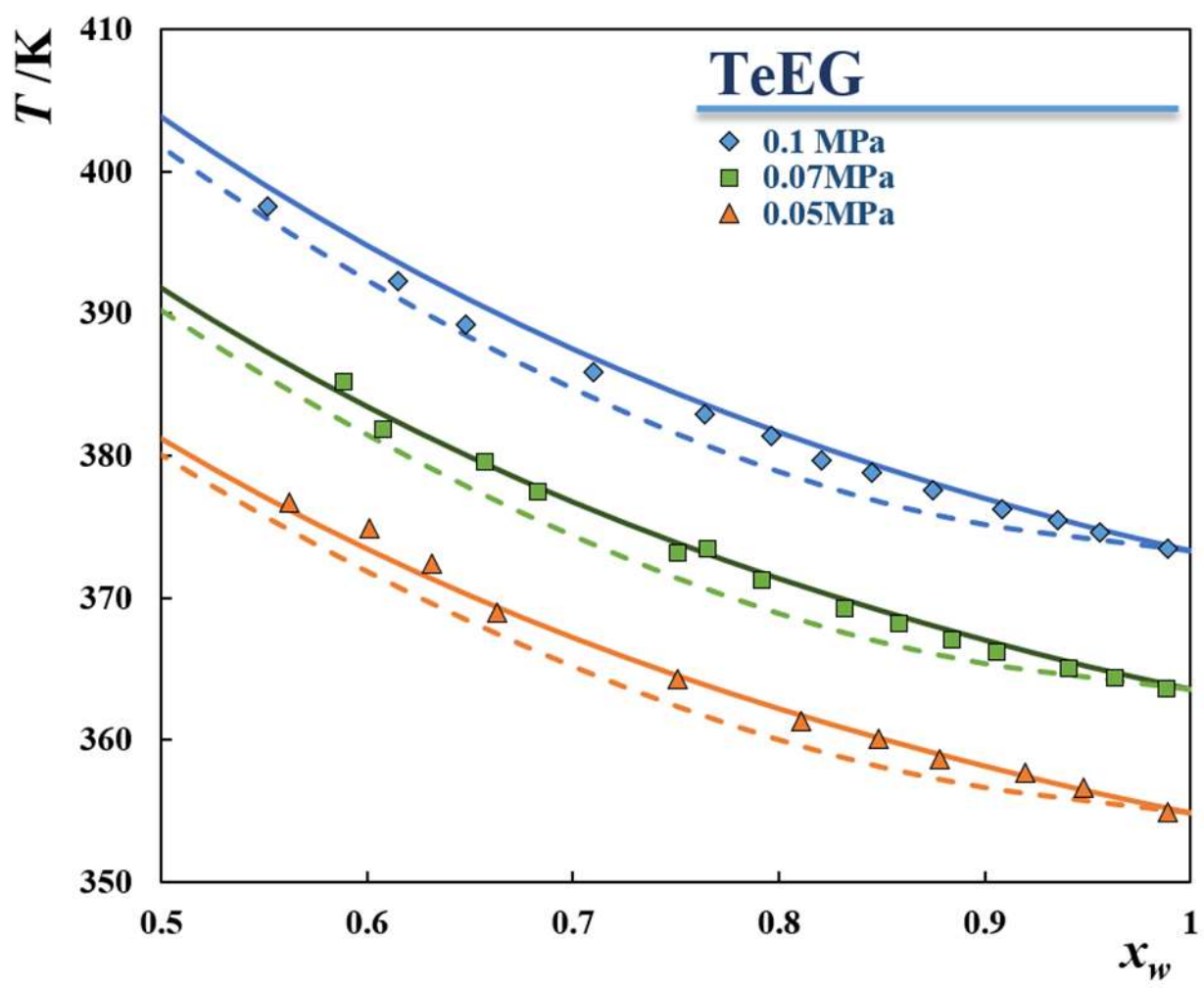

**Figure 10.** Vapor-liquid equilibrium of glycol + water systems. Symbols represent experimental data[82] while the solid and dashed lines represent the SAFT-γ-Mie results using approach A and approach B, respectively.

**Table 10.** Deviations between the modelling results and the VLE experimental data for glycol + water systems.

| | Approach A | Approach B |
|---|---|---|
| System | AAD (T) /K | |
| DEG + water | 0.76 | 5.55 |
| TriEG + water | 0.60 | 0.92 |
| TeEG + water | 0.57 | 1.53 |
| System | % AARD ($a_W$) | |
| PEG 200 + water | 4.28 | 5.96 |
| PEG 400 + water | 3.72 | 16.78 |
| PEG 600 + water | 7.29 | 22.93 |
| PEG 1500 + water | 4.07 | 50.39 |
| PEG 6000 + water | 5.07 | 33.45 |

Different authors[83–85] have reported the water activities in polyethylene glycol aqueous solutions for polymers of different molecular weight and at different temperatures. As the parameters governing the interactions between EO groups and water are already defined, we can use the model to predict the behavior of such systems. As for higher molecular weight polymers the temperature effect on the water activities is negligible, the results of these predictions at 298.15 K are shown in **Figure 11** for PEGs in the molecular weight range 200-6000 g/mol. As depicted in the figure, approach A performs much better than approach B, providing an excellent agreement with the experimental data while approach B not only shows considerable deviations from the experimental data but also shows a different qualitative behavior with the model predictions showing the existence of a maximum on the water activities that is not observed experimentally.

The results obtained for the glycols/PEGs + water systems show that despite the slightly better performance of approach B to describe the properties of pure small glycol oligomers, approach A performs much better when the parameters are used to predict the behavior of mixtures, namely glycols with either *n*-heptane or water. This reinforces the already known importance of including mixture data in the parameterization of SAFT model parameters, or at least their use to guide the selection of the most appropriate parameter set.[80,86]

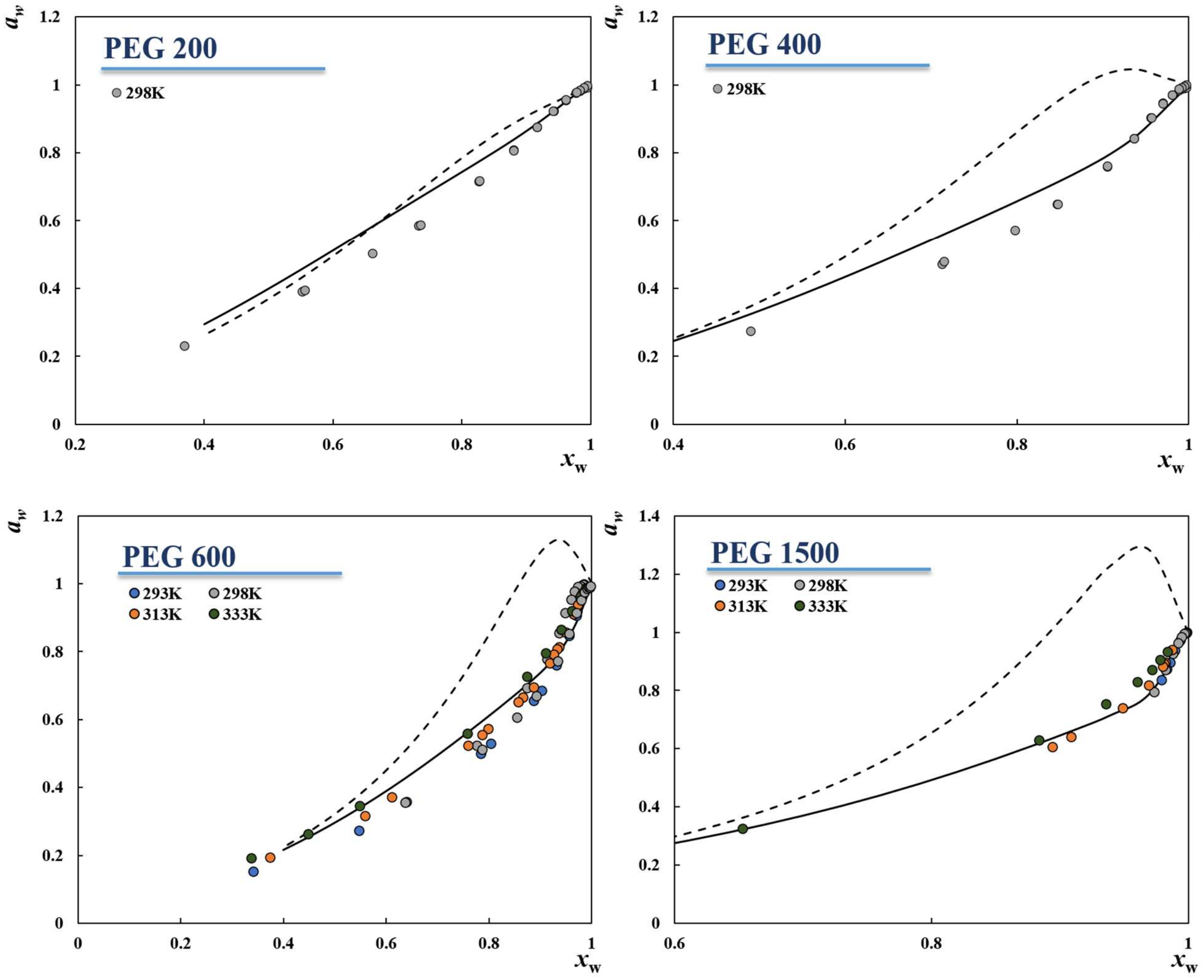

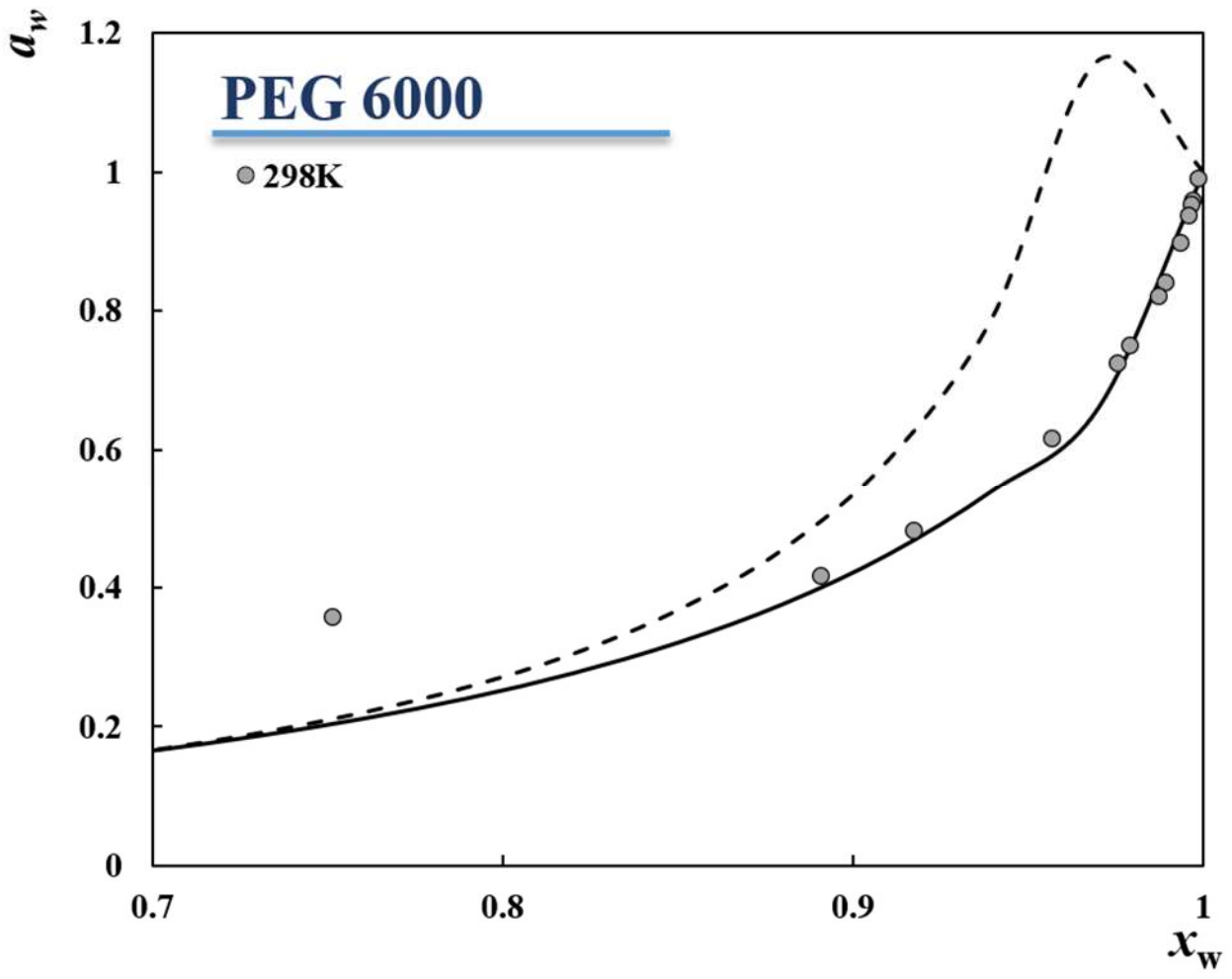

**Figure 11.** Water activity for binary systems water + polyethylene glycols. Symbols represent experimental data[83–85] while the solid and dashed lines represent the SAFT-g-Mie predictions using approach A and B, respectively.

At higher pressures, PEG + water systems exhibit a closed-loop liquid-liquid equilibrium, presenting both a LCST and UCST that results from a delicate balance between the hydrogen bonding and the weaker van der Waals interactions.[87,88] Unfortunately, none of the two approaches was able to predict this phase behavior under a broad range of $(T, p, x)$ conditions tested around the experimental values. As an example, PEG 3350 + water exhibit a closed loop immiscibility between 420-520 K for polymer weight fractions between ~ 0 and 0.5 which corresponds to a polymer mole fraction of less than 0.005. If isothermal flash calculations are carried out in this temperature interval and for a feed containing a polymer mole fraction ranging between 0 and 0.005, the model always predicts a single phase or a single liquid phase with a pure water vapor phase if pressure is lower than the water's vapor pressure at the given temperature.

### 3.8. Pure glymes (glycol ethers)

In this work, the term 'glymes' is used to describe glycol ethers that, contrarily to the typical *CiEj* surfactants, do not present surface activity, mainly due to their very short alkyl chains ($i \leq 3$). This type of components are typically used in industrial processes for the removal of acid gases, an example being the state-of-art Selexol process that typically employs a blend of glycol dimethyl ethers as a physical solvent for the removal of $H_2S$ and $CO_2$ from natural gas streams.[89,90]

The glymes investigated in this work are presented in **Table 11** along with their decomposition into different functional groups. Concerning their chemical structure, these glymes are essentially glycols in which one (glymes mono-alkyl ethers) or both (glycols di-alkyl ethers) terminal hydroxyl groups are replaced by a methyl or ethyl group. Considering the group decomposition for the mono-alkyl ethers, presented in **Table 11**, EGEE, DEGME, DEGEE, and TeEGME can be modelled using the same functional groups previously considered for glycols, without any further fitting.

**Table 11.** Glymes investigated in this work.

| Glyme | Group Decomposition | Chemical structure |
|---|---|---|
| Ethylene glycol ethyl ether (EGEE) M = 90.12 g·mol$^{-1}$ | 1x$CH_3$; 1x$CH_2$; 1xOH; and 1xEO | 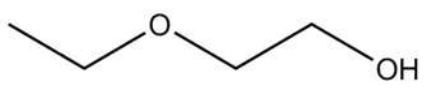 |
| Diethylene glycol methyl ether (DEGME) M = 120.15 g·mol$^{-1}$ | 1x$CH_3$; 1xOH; and 2xEO | 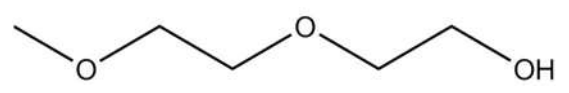 |
| Diethylene glycol ethyl ether (DEGEE) M = 134.17 g·mol$^{-1}$ | 1xCH3; 1xCH2; 1xOH; and 2xEO | 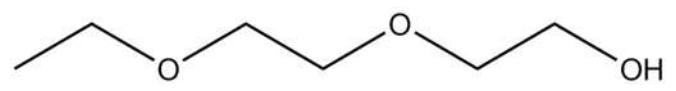 |
| Tetraethylene glycol methyl ether (TeEGME) M = 208.25 g·mol$^{-1}$ | 1x$CH_3$; 1xOH; and 4xEO | 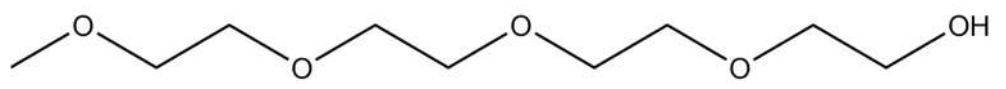 |
| Diethylene glycol dimethyl ether (DEGDME) M = 134.17 g·mol$^{-1}$ | 1x$CH_3$; 2xEO; and 1x$CH_3O$ | 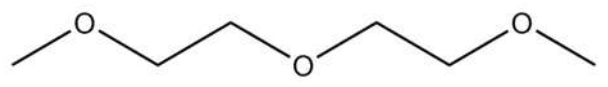 |
| Triethylene glycol dimethyl ether (TriEGDME) M = 178.23 g·mol$^{-1}$ | 1x$CH_3$; 3xEO; and 1x$CH_3O$ | 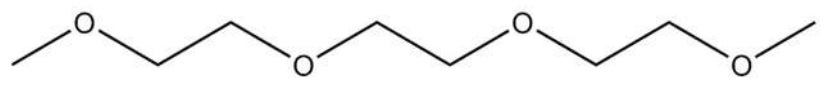 |
| Tetraethylene glycol dimethyl ether (TeEGDME) M = 222.28 g·mol$^{-1}$ | 1x$CH_3$; 4xEO; and 1x$CH_3O$ | 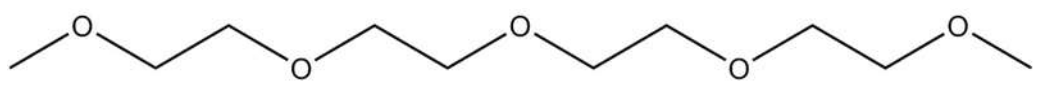 |

Therefore, the group and unlike interaction parameters already defined in the previous sections, using the 'EOa' for the EO group, were used to predict the vapor pressures and saturation liquid densities of the different glymes mono-alkyl ethers, and these are depicted in **Figure 12a** and **Figure 13a**, respectively. As can be observed, the transferred parameters are unable to properly describe the thermodynamic behavior of pure glycol ethers showing very high deviations from the experimental data, as reported in **Table 12**. Additional attempts using the modelling approach B presented in the previous section also resulted in a similar or even lower performance. To improve such results, and to have a set of parameters that adequately describes glymes (and that, consequently may, or may not, be appropriate for the modelling of *CiEj* surfactants at a later stage), a reparameterization of the energy-related parameters involving the EO group, 'EOg' – approach G, and their unlike interactions with $CH_3$, $CH_2$ and OH, using the VLE experimental data of these glycol ethers, was carried out and the new parameters are reported in **Tables 1-2**. The results of this fitting are shown in **Figure 12b** and **Figure 13b** and, as can be observed, although there is a considerable improvement of the results, the deviations from the experimental data are still considerable (cf. **Table 12**).

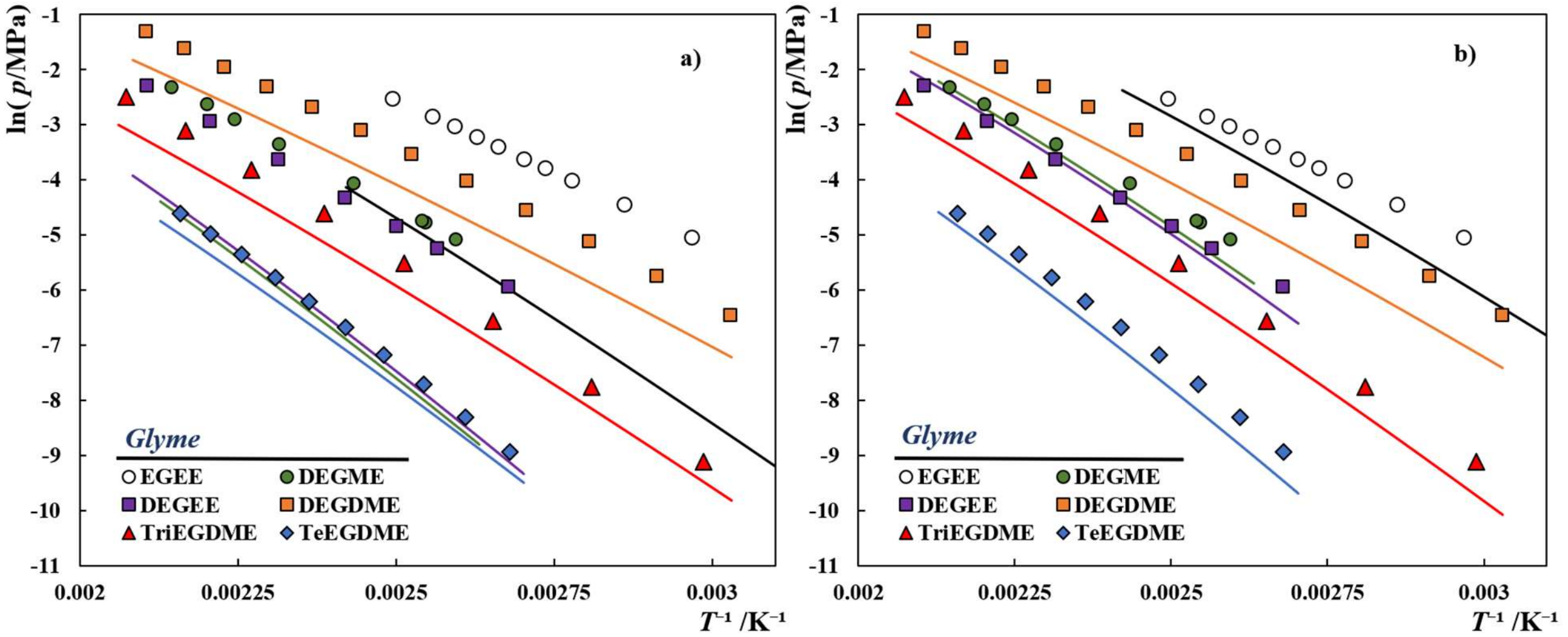

**Figure 12.** Vapor pressures of pure glymes. Symbols represent experimental data[49] while the solid lines depict the SAFT-γ-Mie EoS results using a) approach A ; b) approach G.

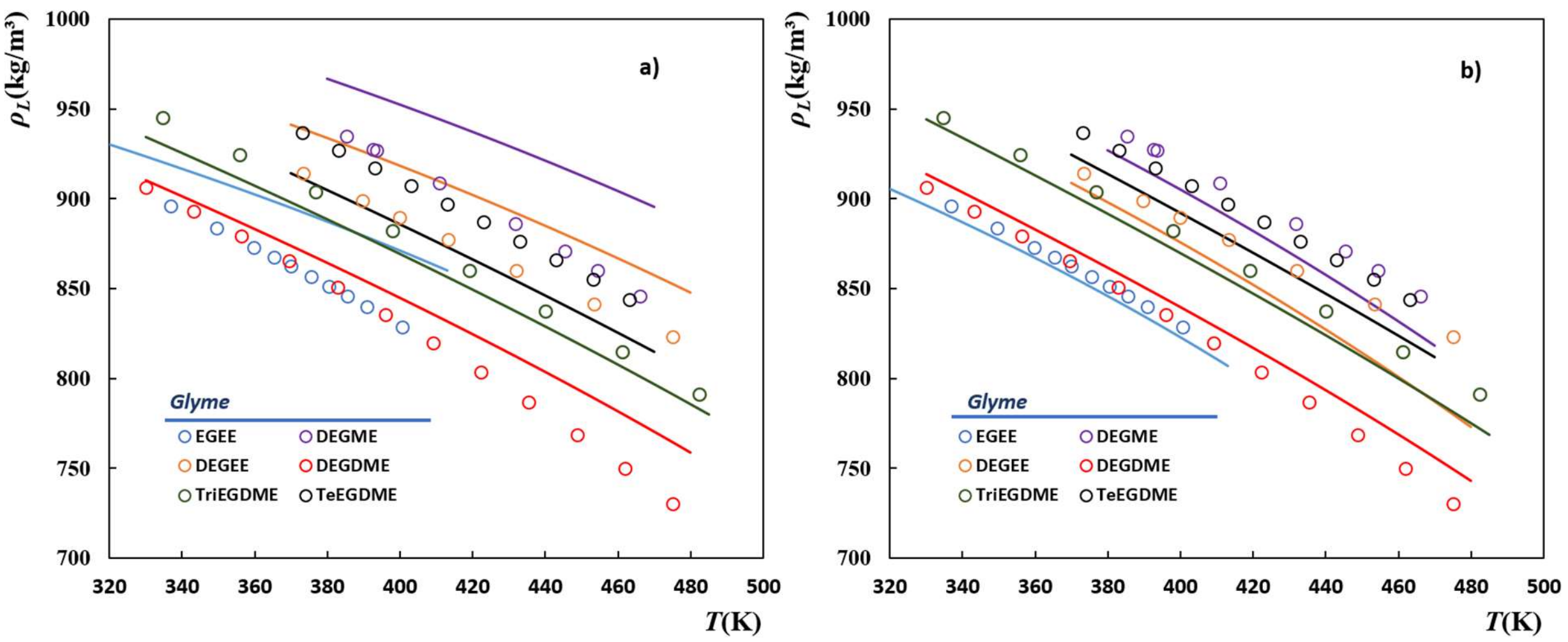

**Figure 13.** Saturation liquid densities of pure glymes. Symbols represent experimental data[49] while the solid lines depict the SAFT-γ-Mie EoS results using a) approach A ; b) approach G.

**Table 12.** Deviations from the experimental VLE data of pure glymes, reported in %AARD.

| | Approach A | | | Approach G | | |
|---|---|---|---|---|---|---|
| | $\rho_L$ | $p^*$ | $p\rho T$ | $\rho_L$ | $p^*$ | $p\rho T$ |
| **EGEE** | 3.87 | 91.79 | 1.25 | 0.67 | 39.20 | 1.04 |
| **DEGME** | 4.45 | 93.46 | 0.72 | 1.94 | 21.36 | 1.25 |
| **DEGEE** | 3.38 | 90.32 | 0.88 | 2.48 | 14.68 | 0.74 |
| **TeEGME** | - | - | 0.90 | - | - | 1.14 |
| **DEGDME** | 2.06 | 50.30 | 0.44 | 1.35 | 49.13 | 0.60 |
| **TriEGDME** | 1.21 | 38.74 | 1.18 | 1.36 | 36.08 | 0.41 |
| **TeEGDME** | 2.64 | 32.65 | 2.08 | 2.22 | 31.31 | 0.79 |

The very low accuracy of the model, when used to describe pure glycol ethers, suggests that the effect of some of the functional groups composing the mono-alkyl glymes are not well captured by the model, even though the 'EOg' group has been parameterized to pure glymes VLE data. This is in good agreement with the lack of transferability previously observed for other groups, as it is the case with the hydroxyl group, whose transferability from alkan-1-ols to α, ω – alkanediols and glycols seems to be very limited. In the previous sections, this could be partially explained by the additive effect of having multiple OH groups in the same molecule as both α, ω – alkanediols and glycols have two terminal hydroxyl groups but this is not the case for the mono-alkyl glymes studied here, where only one terminal hydroxyl group (as in alkan-1-ols) is present and still, huge deviations from the experimental data are observed.

On the other hand, according to the group decomposition proposed in **Table 11**, the study of di-alkyl glymes requires the introduction of a new $CH_3O$ group. The parameters characterizing this new group and their unlike interactions with $CH_3$ and EO groups also present in di-alkyl glymes were obtained by fitting to the experimental vapor pressures and saturated liquid densities of DEGDME, TriEGDME, and TeEGDME available in the DIPPR database,[49] and the final values of the parameters are reported in **Tables 1-2** as 'CH3Oa' and 'CH3Og' groups whether the EO group was modelled following approach A or the new approach G. The results of the fitting are also shown in **Figures 12-13** and the deviations from the experimental data are reported in **Table 12**. Following approach A, as a parameterization of the new group is involved, the accuracy on the description of di-alkyl glymes is better than for mono-alkyl glymes, especially in the vapor pressures, although high deviations are still observed, reinforcing the poor transferability of the remaining functional groups. Concerning, approach G, a similar accuracy was obtained for both types of glycol ethers (in both cases, a parameterization step was taken), with densities being better described for di-alkyl glymes and vapor pressures for the mono-alkyl glymes.

Having available experimental data for the high-pressure liquid densities of different glycol ethers,[10] both modelling approaches (A and G) were used to predict the density in a wide temperature (283-363 K) and pressure (0.1-95 MPa) ranges and the results are shown in **Figures S7-S8**, in Supporting Information. The deviations from the experimental data provided in **Table 12** show that despite the extra parameterization step in approach G, the accuracy of both approaches in predicting the $p\rho T$ data is similar, reinforcing the idea that the extra parameterization does not solve the issues associated with the transferred groups, e.g. OH. Nonetheless, the deviations are lower than those observed for the saturation liquid densities, suggesting that the models perform much better at lower temperatures, far from the saturation conditions of the pure component.

### 3.9. Glymes + water systems

In a recent work,[91] the experimental boiling temperatures of aqueous solutions of six different glycol ethers, all of them studied in the previous section, were reported at three different pressures (0.10, 0.07, and 0.05 MPa). Hence, in this section, the existent coarse-grained water model is used along with the two glymes modelling approaches (A and G), discussed previously, to describe the available experimental data.

For mono-alkyl glymes, using approach A, all the necessary parameters are already available, since the unlike interactions between EO and $H_2O$ groups were already characterized while studying glycols + water mixtures. On the other hand, for approach G, the required $\varepsilon_{EO-H2O}$, $\varepsilon^{HB}_{EO-H2O,e2-H}$, and $\kappa^{HB}_{EO-H2O,e2-H}$ parameters have to be fitted to the experimental data at 0.1 MPa for EGEE, DEGME, and DEGEE aqueous solutions and used to predict the behavior at the two other pressures investigated. For di-alkyl glymes, as a new $CH_3O$ group was introduced, the parameters characterizing the unlike interactions between this group and water are necessary in both approaches, and were fitted to the experimental VLE data of aqueous solutions of DEGDME, TriEGDME, and TeEGDME at 0.1 MPa.

The modelling results are shown in **Figure 14**, and the deviations from the experimental data, reported as an average absolute deviation in K, are provided in **Table 13**.

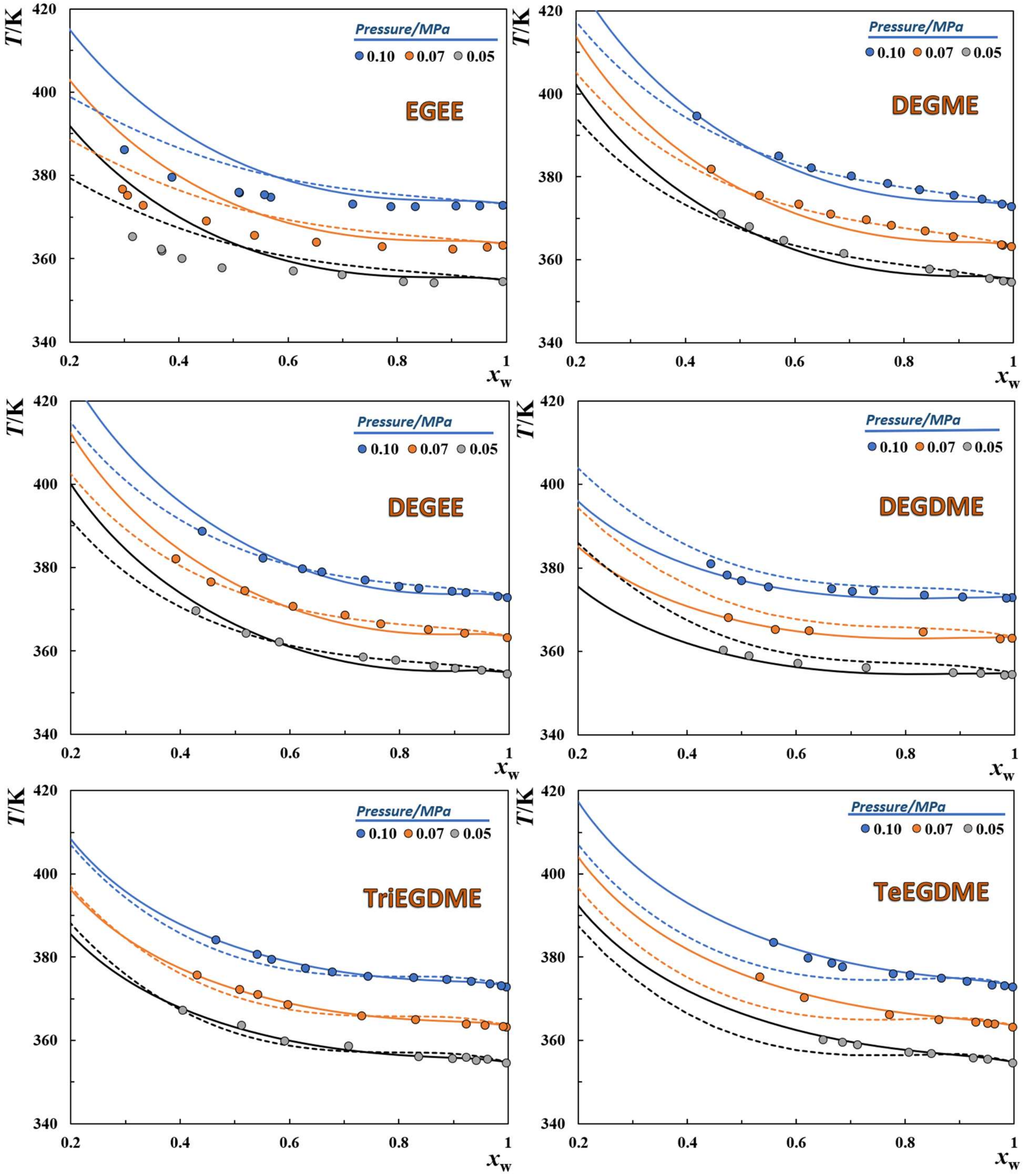

**Figure 14.** Boiling temperatures of glymes + water mixtures at different pressures. Symbols represent experimental data[91] while the solid and dashed lines represent the SAFT-γ-Mie results using approach A and approach G, respectively.

**Table 13.** Deviations between the modelling results and the VLE experimental data for glyme + water systems.

| | Approach A | Approach G |
|---|---|---|
| System | AAD (T) /K | |
| EGEE + water | 5.6 | 4.14 |
| DEGME + water | 1.38 | 0.49 |
| DEGEE + water | 1.11 | 0.55 |
| DEGDME + water | 0.63 | 2.08 |
| TriEGDME + water | 0.37 | 1.06 |
| TeEGDME + water | 0.64 | 1.90 |

As expected, for the mono-alkyl glymes, approach G performs slightly better than approach A, since the unlike interactions between 'EOg' and water were specifically parameterized for these systems. Nevertheless, the results obtained using approach A suggest the robustness of this modelling approach when used to describe the VLE of aqueous solutions of EO-containing compounds, as previously seen for glycol + water mixtures. For the di-alkyl glymes, where the $CH_3O$-water interaction parameters were fitted in both cases, approach A provides the most accurate results, reinforcing the overall robustness of approach A.

The good performance of approach A, when used to describe the VLE of aqueous solutions of both glycols and glymes, is somehow surprising considering that this modelling approach presented considerable deviations from the experimental vapor pressures of the pure glycols/glymes, for which approach B and approach G yielded the best results, respectively. Evidently, in the VLE of such aqueous solutions the contribution of the polyether to the total pressure is very small. Therefore, given that the water model is the same in all the calculations, the accuracy of the experimental vapor pressure data available for pure polyethers may be questioned.

This reinforces the idea that mixture data should indeed be included in the parameterization datasets in order to select and validate the most appropriate set of parameters for a given group or group pair as, clearly, a better representation of pure-component properties do not necessarily result in a better description of mixtures behavior, and there are several sets of parameters yielding results with a similar accuracy in what concerns pure-component properties.

### 3.10. *CiEj* non-ionic surfactants in water

In this section, the performance of the three modelling approaches previously suggested for glycols and glymes (A, B, and G) when used to describe the behavior of *CiEj* non-ionic surfactants and their aqueous solutions is evaluated, since these surfactants are only composed of $CH_3$, $CH_2$, EO, and OH groups, all of them extensively analyzed in the previous sections.

$C_4E_1$ is the shortest of this type of compounds that exhibit measurable surface activity and due to its low melting point, one of the few for which pure fluid experimental data is available. Therefore, the three modelling approaches were used to predict the liquid densities and vapor pressures of $C_4E_1$, and the results are shown in **Figure 15** below.

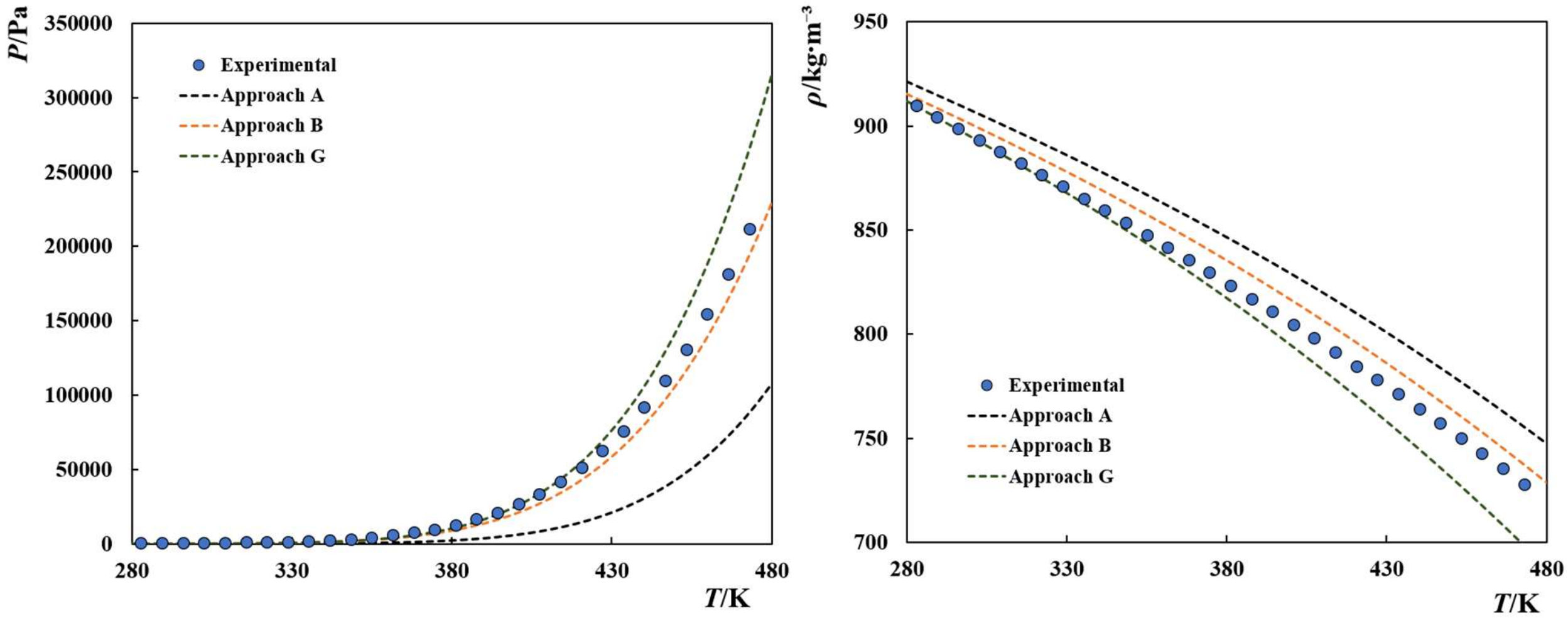

**Figure 15.** Vapor pressures and liquid densities of pure $C_4E_1$. Symbols represent experimental data[49] while the dashed lines represent the modelling predictions using different approaches.

From the figure, one can observe that while the modelling approach A clearly fails to describe both properties, the modelling approach B provides a reasonable prediction of both properties, slightly overpredicting the liquid density while underpredicting the vapor pressures. On the other hand, approach G seems to provide a good description of the experimental data at low temperatures, while presenting similar deviations to those exhibited by approach A at higher temperatures.

Concerning the aqueous solutions of *CiEj* surfactants, one of the main features is the presence of an immiscibility region in their phase diagram. Usually, at very low temperatures, the system consists of a homogeneous solution but, with increasing temperature, a phase separation occurs above a lower critical solution temperature (LCST), into a water-rich phase and a surfactant-rich phase. This immiscibility can persist at higher temperatures, although in many cases it disappears above an upper critical solution temperature (UCST), resulting in a closed-loop LLE, which is commonly observed in *CiEj* + water mixtures.

This phenomenon results from a delicate balance between energetic and entropic contributions to the systems energy with the magnitude of both the van der Waals and hydrogen bond interactions playing a crucial role. If at very high temperatures, the system maximizes its entropy by being homogeneous, the presence of unfavorable van der Waals forces between water and the surfactant induce phase separation as the temperature is decreased below the UCST. However, if the temperature is further decreased (to temperatures lower than the LCST) the enthalpic contribution due to the very strong attractive hydrogen bonding interactions between the two species overcomes the other effects, resulting in the formation of a single phase. Obviously, capturing such delicate balances between entropic/energetic contributions and the magnitude of vdW/HB forces is a challenging task to any thermodynamic model, and represent a stringent test to the transferability of the SAFT-γ-Mie EoS parameters obtained in the previous sections.

Therefore, the three modelling approaches were used to predict the phase behavior of four surfactants, namely $C_4E_1$, $C_6E_2$, $C_8E_4$, and $C_{10}E_4$, in water, and the results are depicted in **Figure 16**. From these systems, the experimental data reported for aqueous solutions of $C_6E_2$ and $C_8E_4$ do not exhibit the presence of a UCST. Often this is because the systems were not investigated at sufficiently high temperatures due to the temperature instability of the molecules.

The modelling results in **Figure 16** show that none of the three approaches can adequately describe any of the selected systems. As an example, none of the modelling approaches predicted the occurrence of a LCST above the freezing temperature of water (273 K). Moreover, except for the modelling approach G on the $C_4E_1$ system, the models do not predict the occurrence of a UCST in a reasonable temperature range but rather a transition from a LLE to a VLE at higher temperatures. Also clear from **Figure 16** is that, except for the $C_4E_1$ system for which approach G seems to provide the most reasonable description of the system, approach A presents the best results for the remaining systems by predicting a narrower immiscibility gap than those predicted by the remaining approaches. This is in agreement with the results previously obtained for glycols and glymes, where approach A, despite having the worst performance when used to describe the pure EO-containing compound, seems to have the best performance when used to model mixtures.

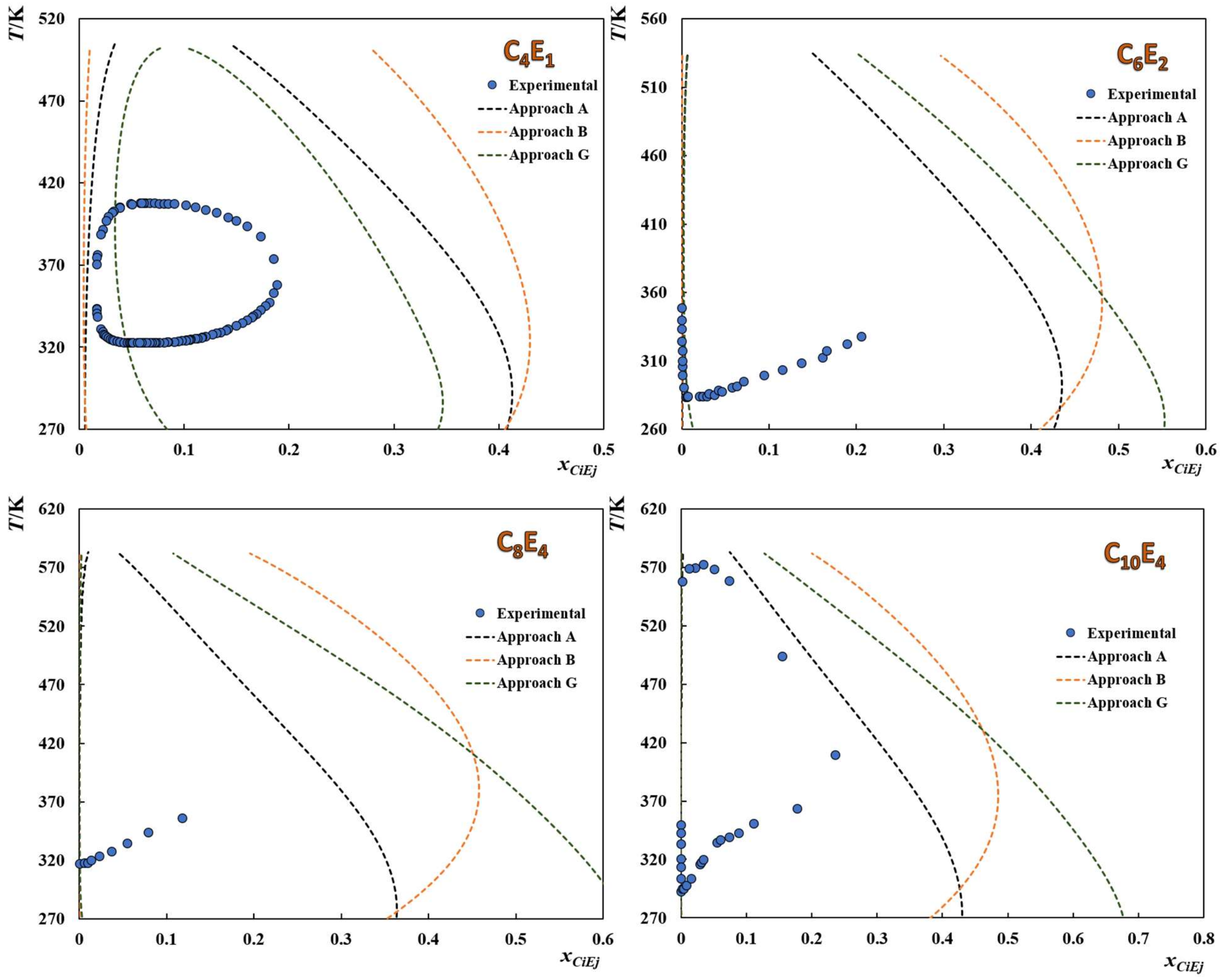

**Figure 16.** Liquid-liquid equilibrium of $C_iE_j$ + water systems. Symbols represent experimental data[92–96] while the dashed lines represent the modelling predictions using different approaches.

As the model parameters transferred from the compounds studied in the previous sections did not yield a closed-loop type LLE for the selected systems, further calculations were carried out by refitting the unlike interaction parameters between the EO group and the water molecules, namely $\varepsilon_{EO-H2}$ , $\varepsilon^{HB}_{EO-H2O,e2-H}$, and $\kappa^{HB}_{EO-H2O,e2-H}$, specifically to closed-loop LLE data of *CiEj* + water systems. The results of such calculations are shown below in **Figure 17** for two different scenarios: represented with a red line are the results of the model when the above mentioned parameters are fitted to the LLE data of $C_4E_1$ + water system while a blue line represents the results of the model when those same parameters are fitted to the LLE data of $C_{10}E_4$ + water system.

The results obtained show that, as previously discussed, whenever a parameterization is carried while studying a new family of compounds, good results, at least qualitatively, can be obtained for most systems. In this case, when the unlike interaction parameters between EO groups and water are fitted to LLE experimental data of a representative *CiEj* + water system, their characteristic closed-loop immiscibility gap is always captured by the model, contrarily to the pure predictions shown in **Figure 16**.

However, even by refitting these parameters using data for a representative system, the agreement with the experimental data is still far from perfect (this is partially due to the limitations of the remaining transferred parameters, as discussed previously), and the new parameters are not necessarily transferable to other chain lengths of the hydrophobic or hydrophilic moieties of the surfactant, with the performance of the model deteriorating considerably between the different systems. This is shown by the example presented in **Figure 17.** Here, when using parameters fitted to the LLE data of $C_4E_1$ + water system, very high deviations between the model predictions and the experimental data for the $C_8E_4$ + water system are obtained while, when using parameters that were regressed from LLE experimental data for $C_{10}E_4$ + water system, the critical temperatures of the $C_4E_1$ + water system are much higher/lower than the experimental UCST/LCST. The results obtained also suggest that a simultaneously good agreement with both the critical temperatures and the two-phase compositions may be difficult to obtain, at least using temperature-independent parameters, as a best description of the phase compositions (blue line) is usually obtained at the expense of higher deviations in the critical temperatures.

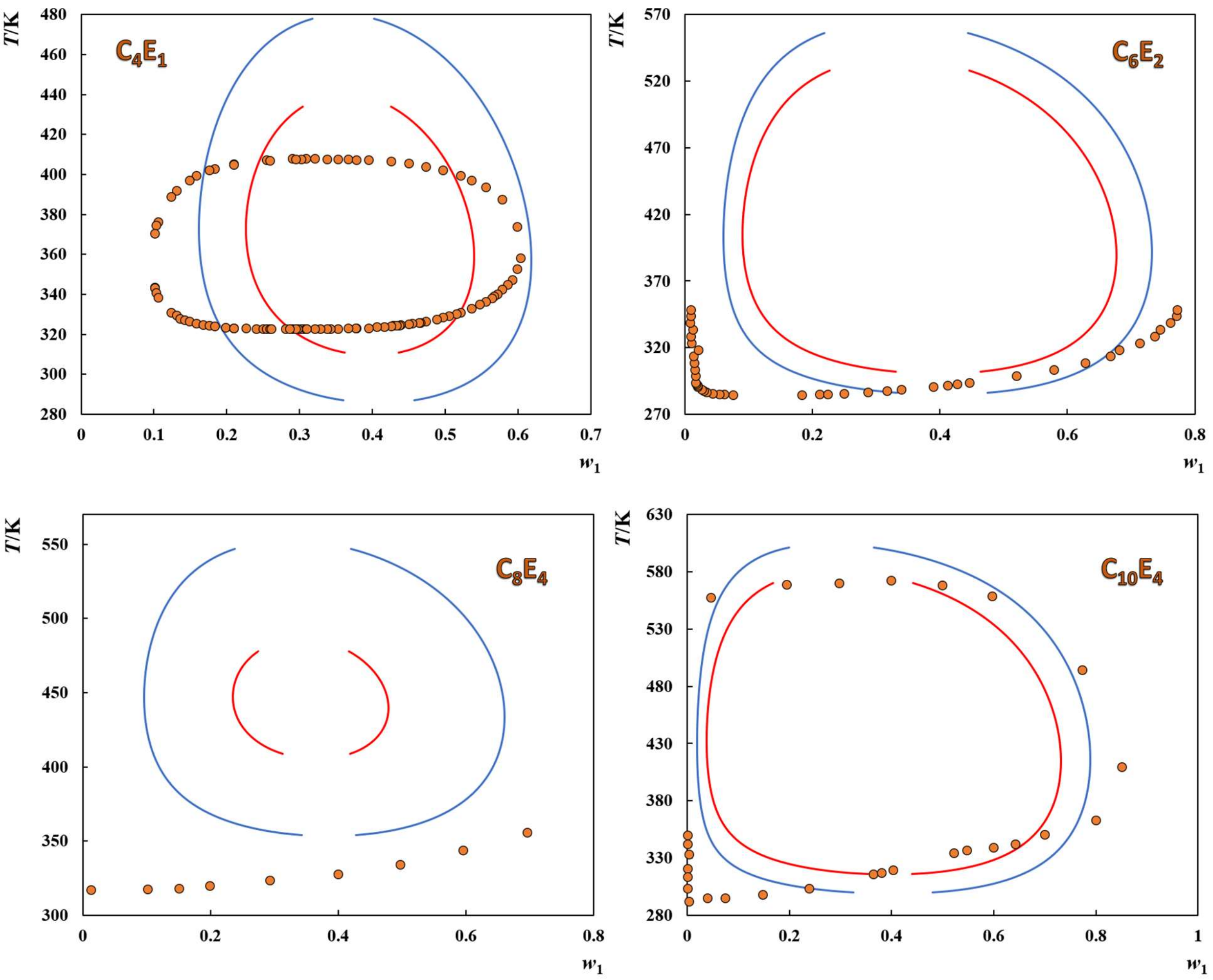

**Figure 17.** Liquid-liquid equilibrium of *CiEj* (1) + water (2) systems. Symbols represent experimental data[92–96] while the red and blue lines represent the SAFT-g-Mie results when the EO-$H_2O$ unlike interaction parameters were fitted to the experimental data for the $C_4E_1$ and $C_{10}E_4$ systems, respectively.

## 4. Conclusions

In this work, a heteronuclear SAFT-type EoS, namely the SAFT-γ-Mie EoS, was applied to the thermodynamic modelling of EO-containing compounds, aiming to evaluate the transferability of the model parameters and if the additional complexity of the heteronuclear treatment of the SAFT theory, leads to an enhancement of the predictive ability and transferability of SAFT models. To accomplish this objective, we have selected three families of EO-containing compounds: glycols, glymes (or glycol ethers), and *CiEj* surfactants, all composed of only $CH_3$, $CH_2$, OH, and EO groups, and thus a suitable choice for tackling the objectives of this study.

Although $CH_3$, $CH_2$, and OH groups are usually transferred from the well-known families of *n*-alkanes and alkan-1-ols and their parameters are usually considered to be good for further applications, this work started by analyzing the performance of such parameters. The results obtained showed that the OH group, typically parameterized using experimental data of alkan-1-ol containing systems, fails to describe other type of components, namely α, ω – alkanediols, for which very high deviations from the experimental data were observed, suggesting that the OH group is not transferable to other families.

As a consequence, when the OH group is used to model glycols, not only the influence of the OH groups in the glycols behavior is not well captured by the model, but the parameterization of the newly introduced EO group is affected, with some of the deviations induced by the problems of the OH group being masked by the EO parameters, perpetuating the problem, with every new addition to the molecular structure.

The results obtained for glycols showed that the model is able to provide reasonably accurate results only when a parameterization is involved, as it is the case of the pure glycol properties or even for the VLE of their aqueous solutions, where the EO-water unlike interaction parameters were parameterized, but fails when used in a purely predictive manner such as in the prediction of the LLE of PEG/water systems, demonstrating the inability of the model parameters to be transferred across different types of phase equilibria.

Afterwards, when the new EO parameters were used to study glymes, the results showed that not only the parameters are not transferable across different types of phase equilibria but they also fail when used across different families of compounds since the model was unable to provide reasonable predictions for pure glymes or their mixtures with water, allowing for a similar conclusion to that reached based on the results obtained for the OH group in the study of α, ω – alkanediols. On the other hand, by reparametrizing the EO group using the glymes experimental data, a much better agreement with the experimental data could be obtained, reinforcing the idea that a (small or large) reparameterization is always present behind the good results reported with heteronuclear SAFT models.

In the final section, the different modelling approaches (those built using glycols experimental data and those based on glymes) were extended for the modelling of *CiEj* surfactants in water and again, the results showed that, without a tweaking of the EO-water and/or OH/water unlike interaction parameters for this type of systems, not even a qualitative agreement with the experimental data could be obtained, with all the modelling approaches failing to predict the existence of a well reported LCST.

The global conclusion of this comprehensive work is that the parameters for the heteronuclear SAFT models are not ready transferable, thus limiting much the predictive capability of this type of models.

**Acknowledgments**

This work was developed within the scope of the project CICECO - Aveiro Institute of Materials, UIDB/50011/2020 & UIDP/50011/2020, financed by national funds through the FCT/MCTES, and when appropriate co-financed by FEDER under the PT2020 Partnership Agreement. This work was funded by PARTEX OIL & GAS. E. A. Crespo acknowledges FCT for the Ph.D. Grant SFRH/BD/130870/2017. We acknowledge Prof. Eduardo J. M. Filipe/Dr. Pedro Morgado, and Lourdes F. Vega for the availability of SAFT codes used for the calculations, and for useful discussions throughout this work.

# Supporting Information

## Insights into the Limitations of Parameter Transferability in Heteronuclear SAFT-type Equations of State

Emanuel A. Crespo,[a] and João A. P. Coutinho[a*]

[a]CICECO – Aveiro Institute of Materials, Department of Chemistry, University of Aveiro, 3810-193 - Aveiro, Portugal;

*Corresponding author: jcoutinho@ua.pt

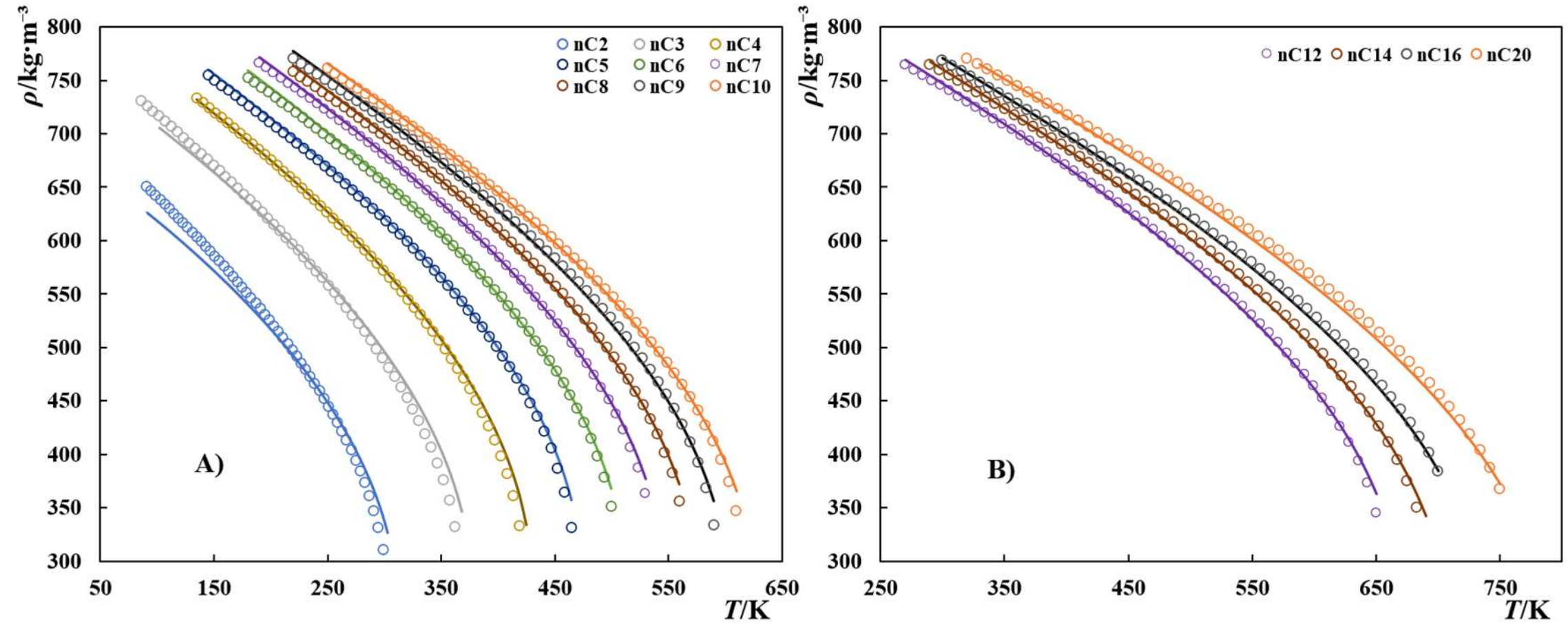

**Figure S18.** Saturated liquid density of linear alkanes. A) from ethane no n-decane (fitting) B) from n-dodecane to n-eicosane (predicted). Symbols represent experimental data from the DIPPR database[1] while the solid lines depict the SAFT-γ-Mie results.

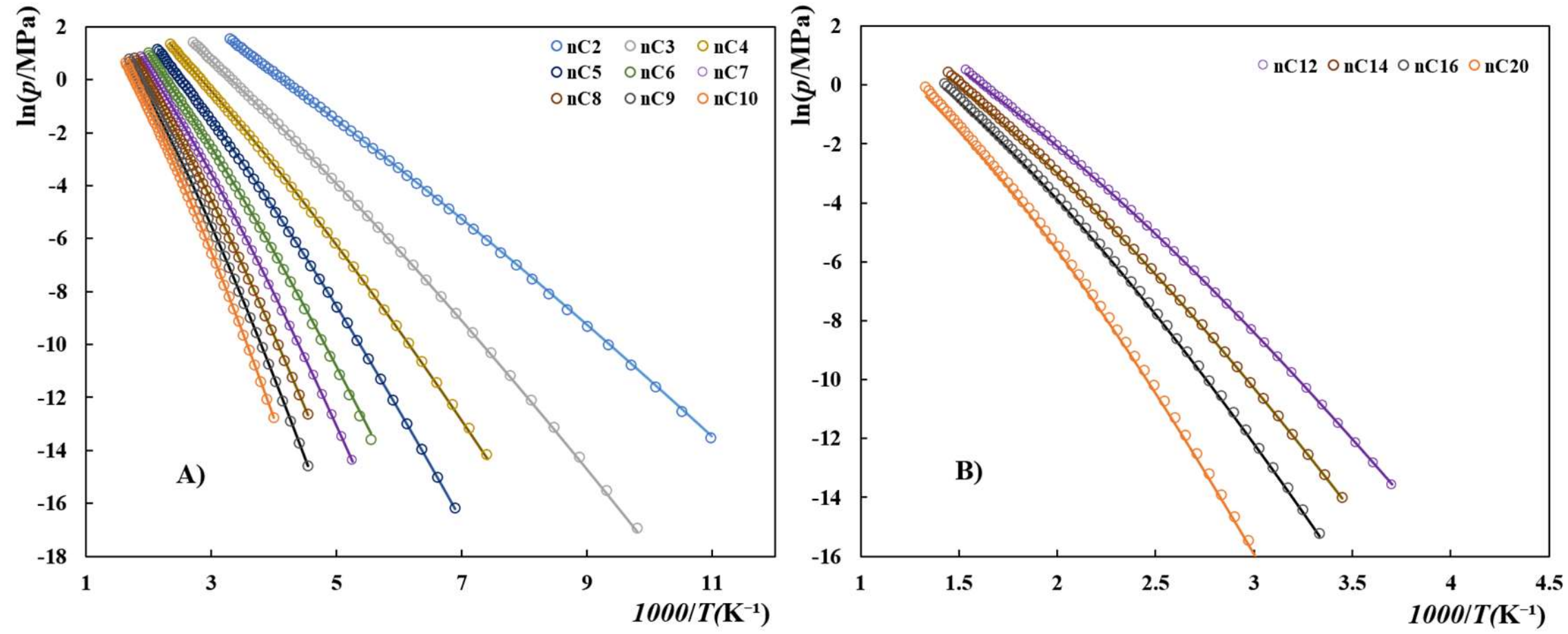

**Figure S19**. Vapor pressures of linear alkanes. A) from ethane to n-decane (fitting) B) from n-dodecane to n-eicosane (predicted). Symbols represent experimental data from the DIPPR database[1] while the solid lines depict the SAFT-γ-Mie results.

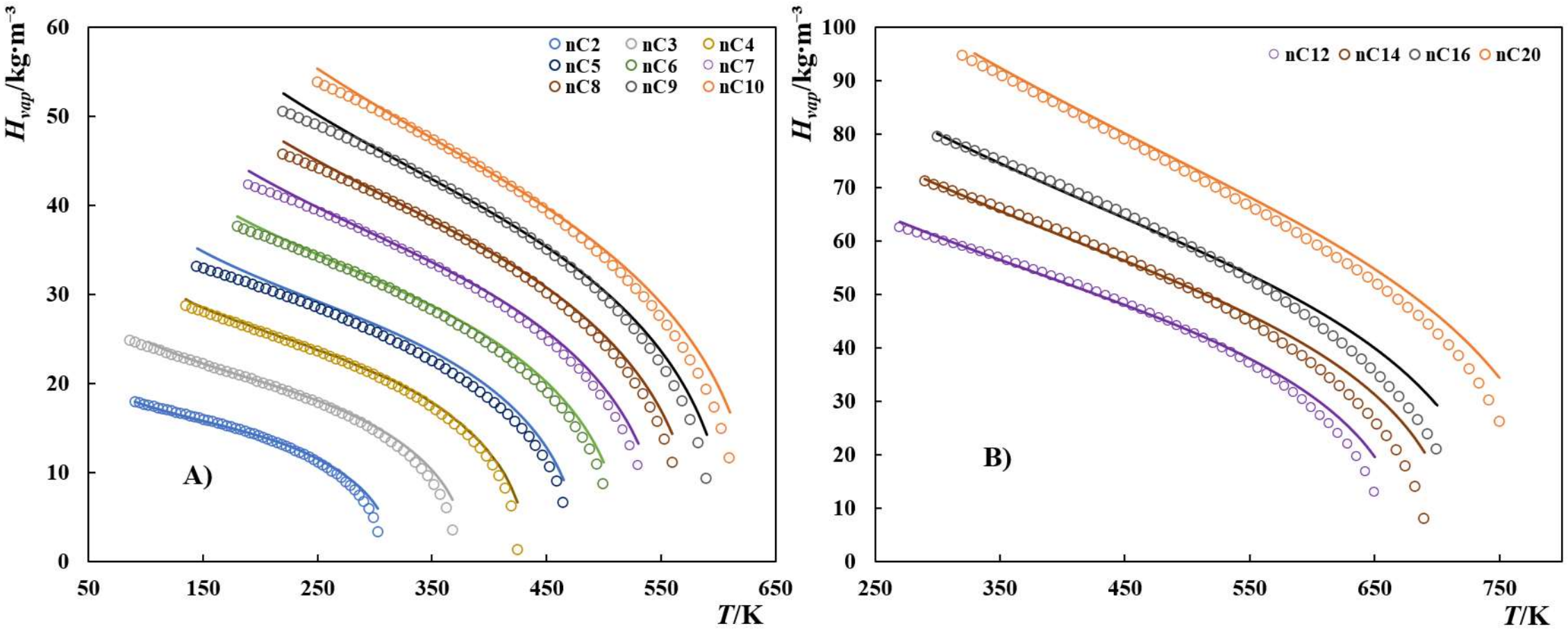

**Figure S20.** Enthalpies of vaporization of linear alkanes. A) from ethane to n-decane (fitting); B) from n-dodecane to n-eicosane (predicted). Symbols represent experimental data from the DIPPR database[1] while the solid lines depict the SAFT-γ-Mie results.

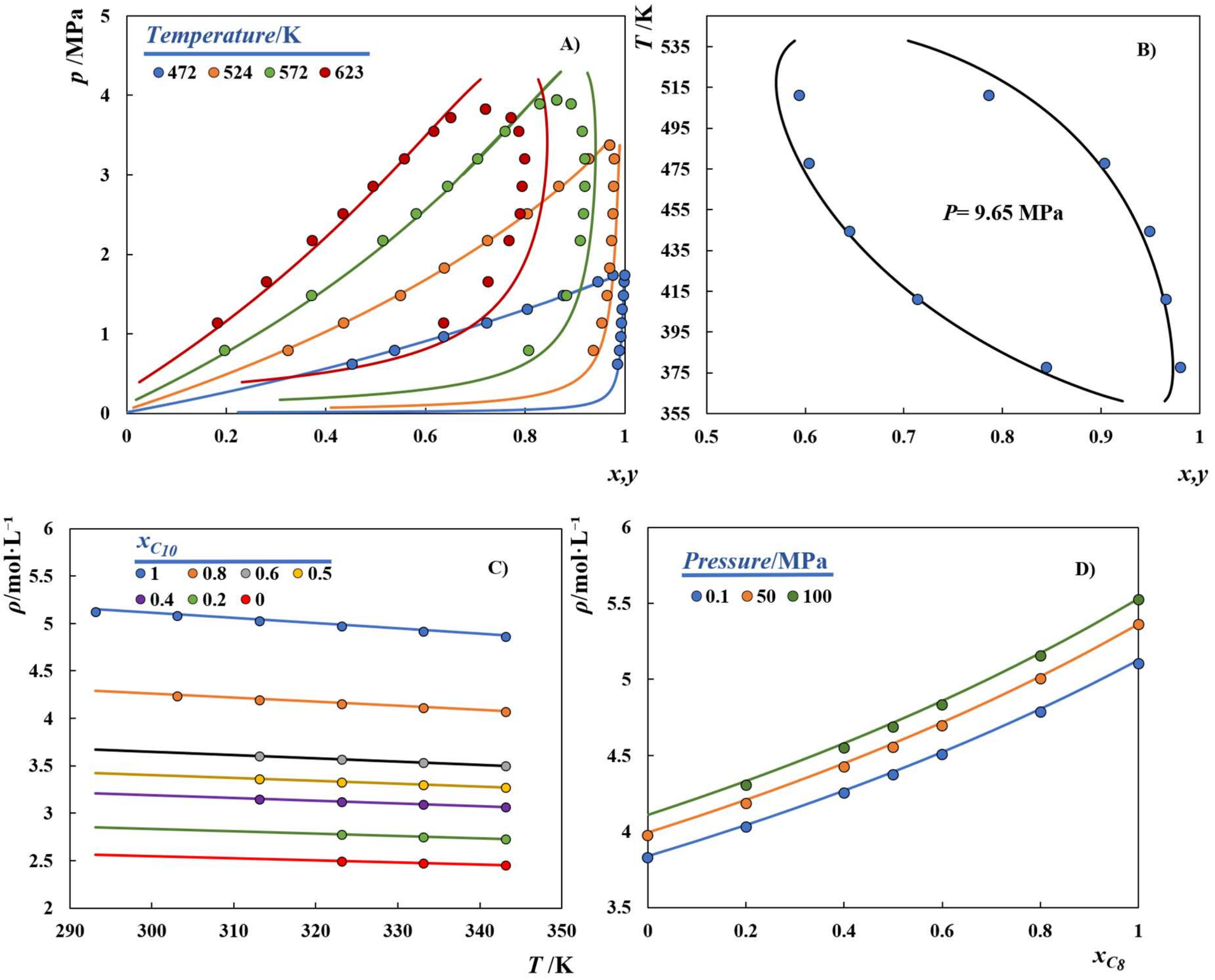

***Figure S21.*** *A) Isobaric VLE of n-hexane + n-hexadecane; B) Isothermal VLE of ethane + n-decane; C) Atmospheric liquid densities of n-decane + n-C22; D) High-pressure liquid densities of n-octane + n-dodecane.*

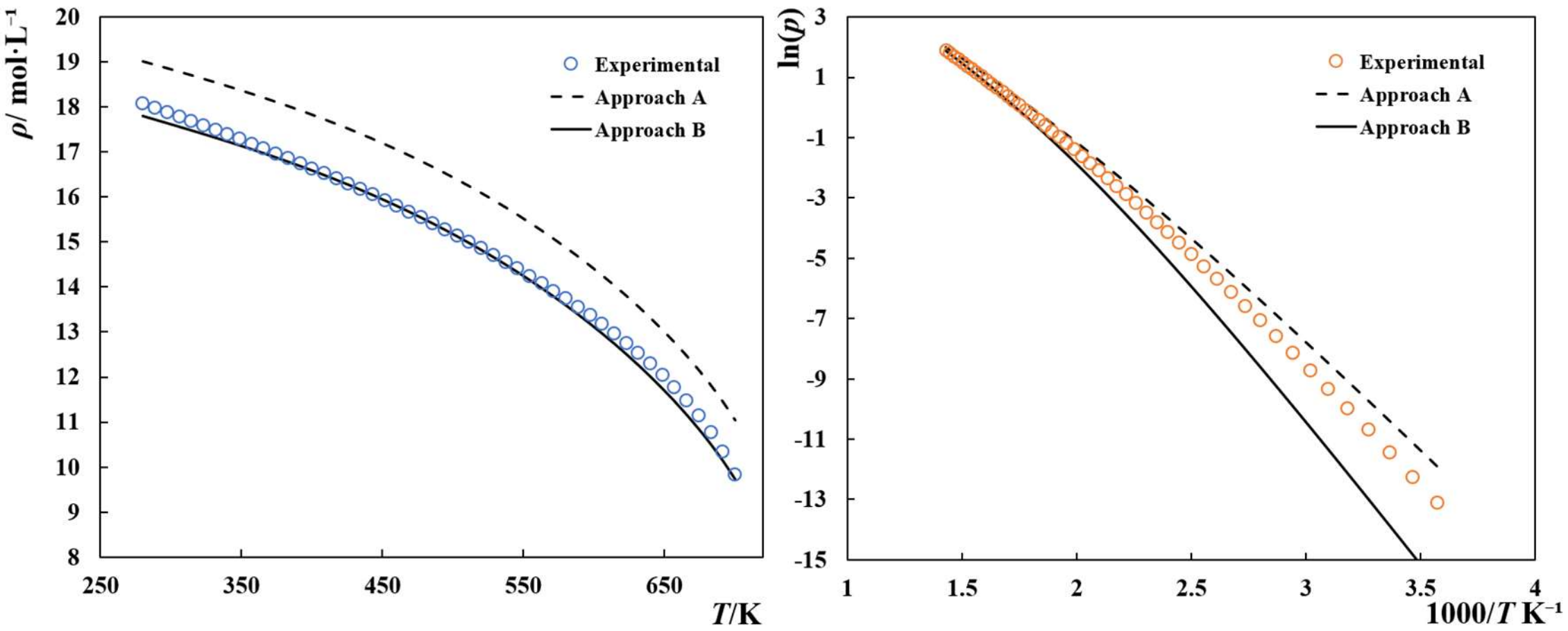

***Figure S22.*** *Saturation liquid densities and vapor pressures of pure ethylene glycol. Symbols represent experimental data*[1] *while the dashed and solid lines represent the SAFT-γ-Mie results following approach A and approach B, respectively.*

5.

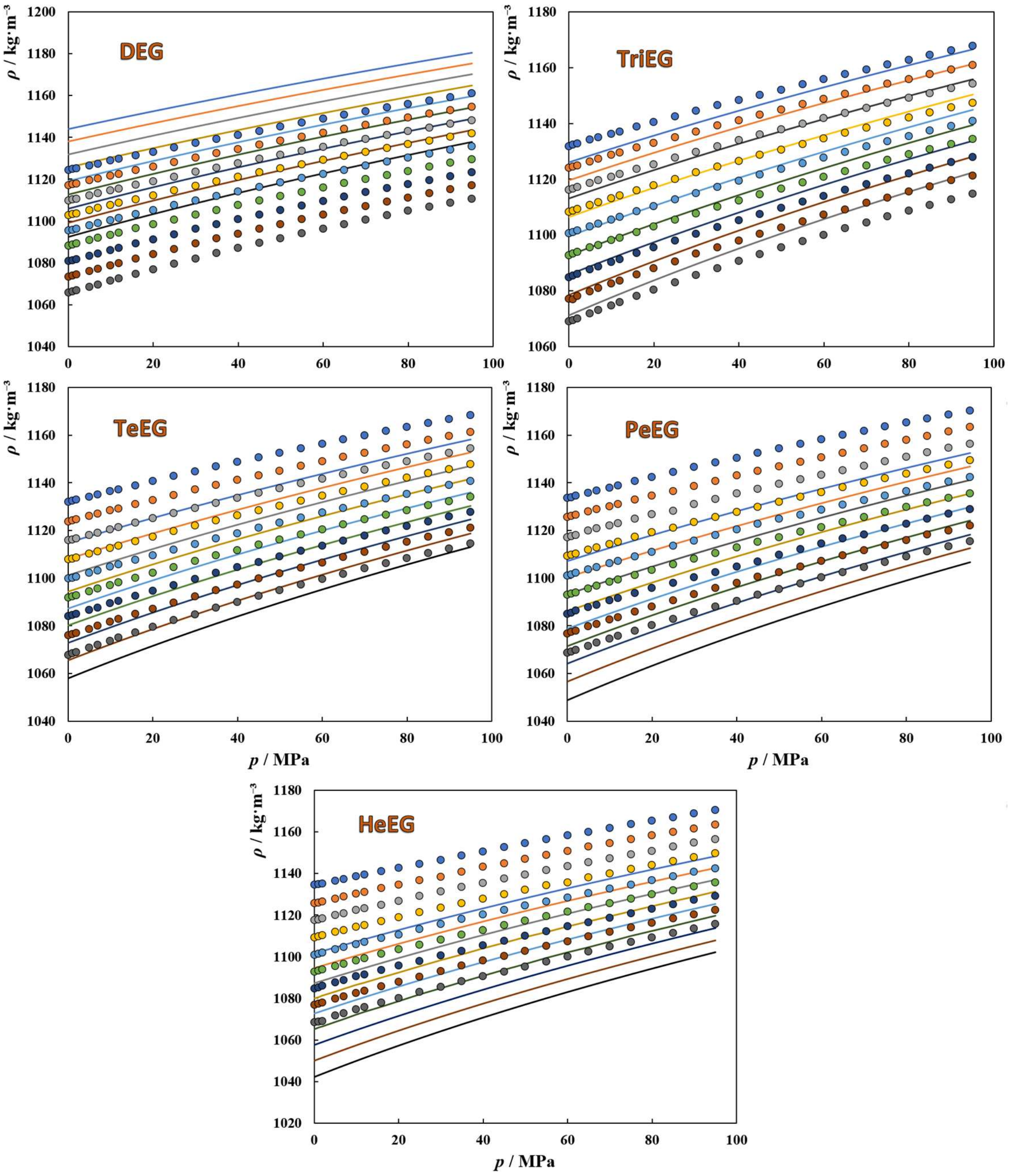

***Figure S23.*** *High-pressure liquid densities of pure glycols. Symbols represent experimental data[2] while the solid lines represent the SAFT-γ-Mie results following approach A.*

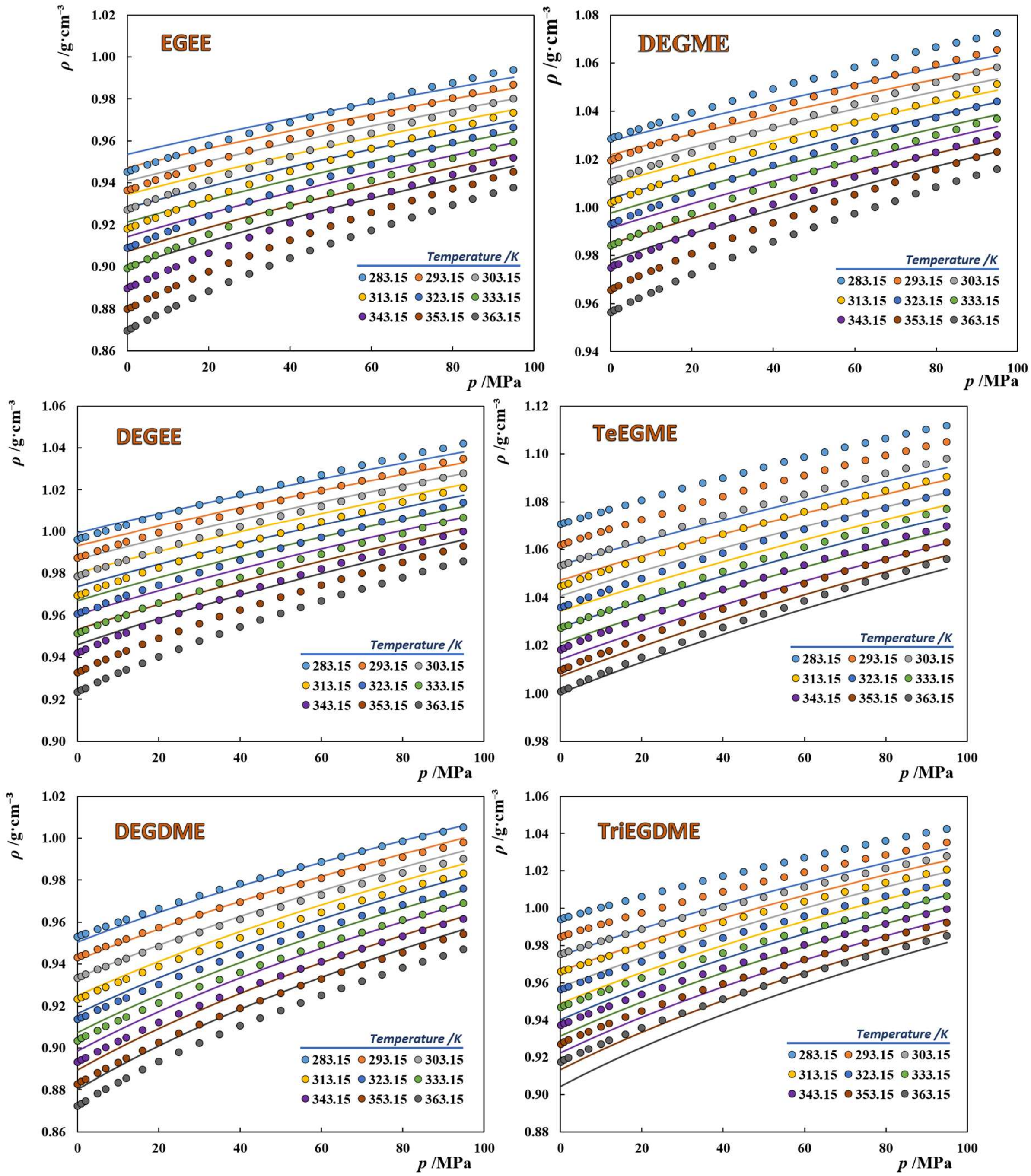

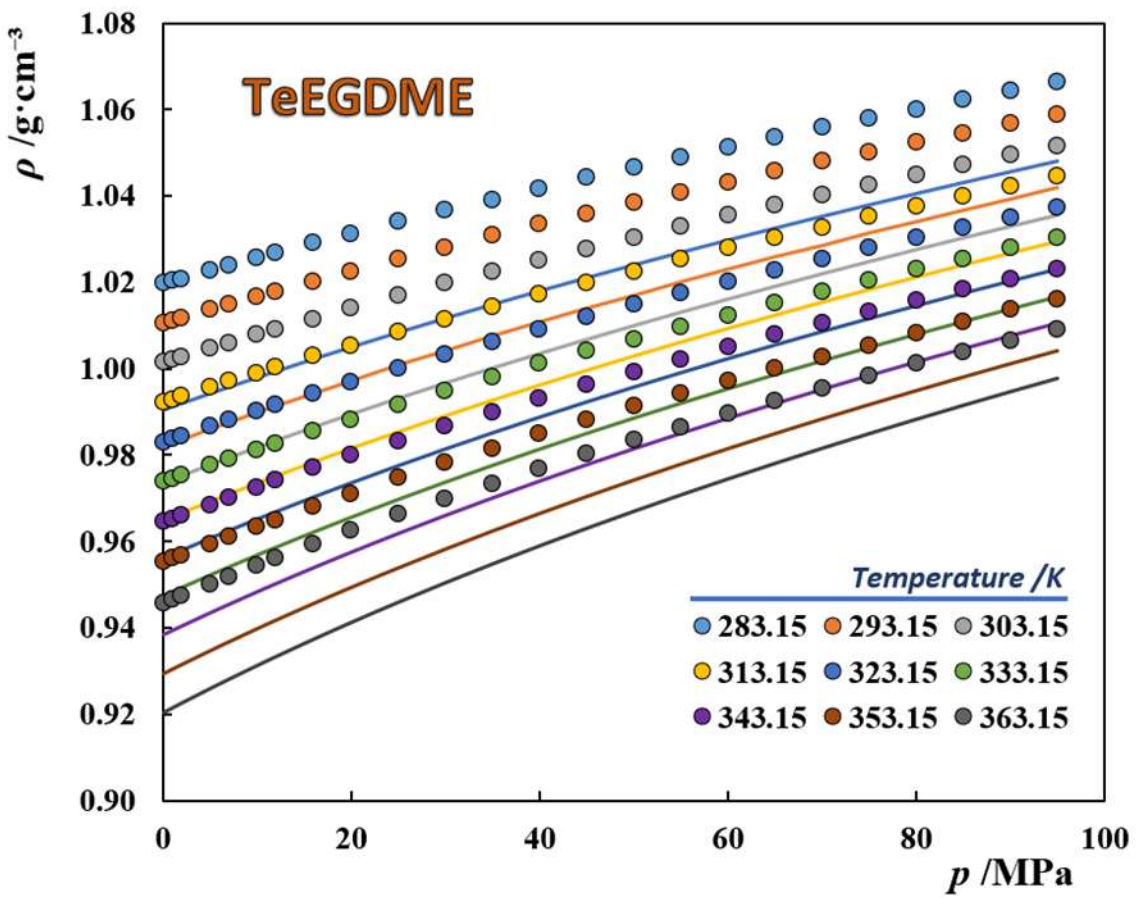

***Figure S24.*** *High-pressure liquid densities of pure glymes. Symbols represent experimental data[3] while the solid lines depict the SAFT-γ-Mie results, following approach A.*

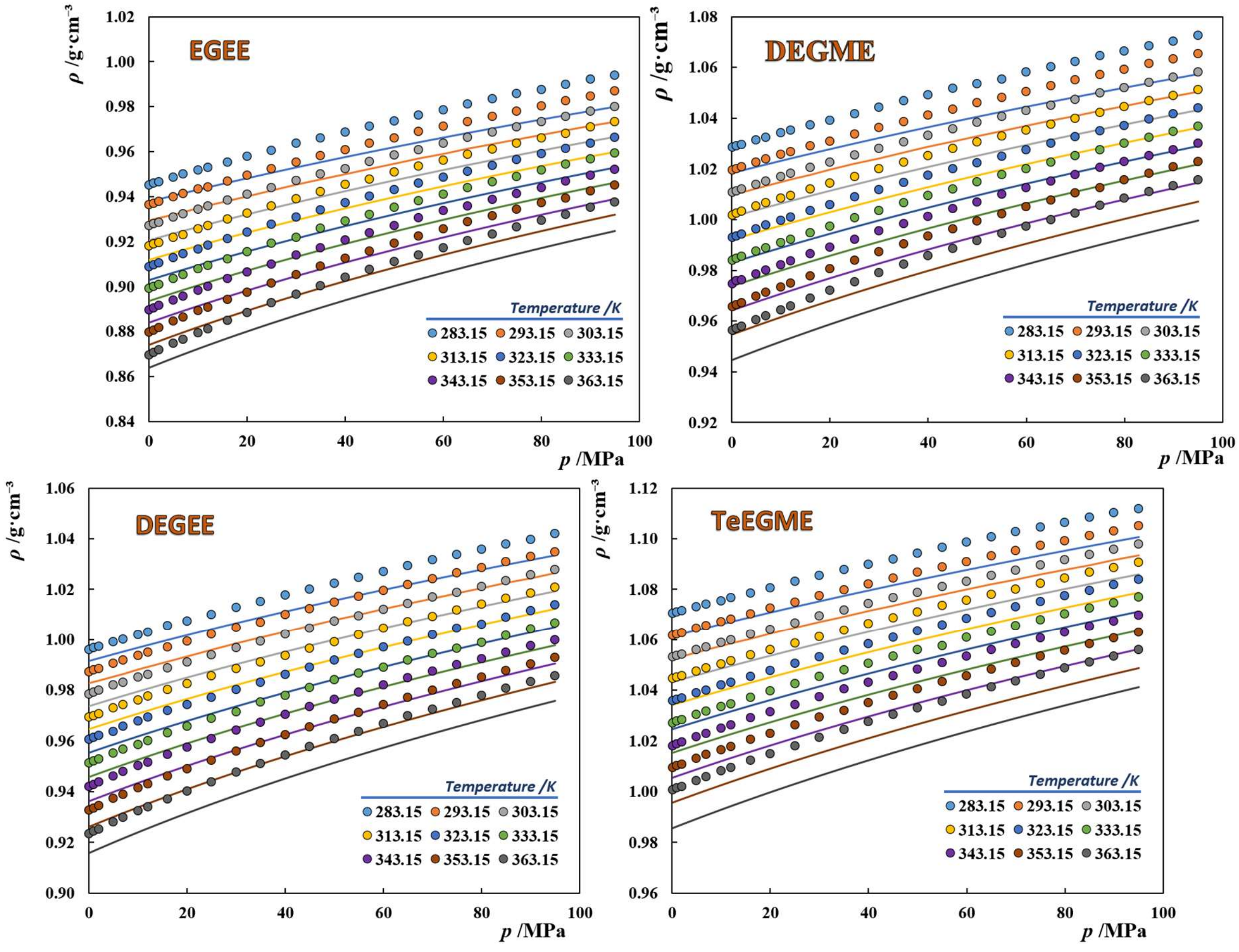

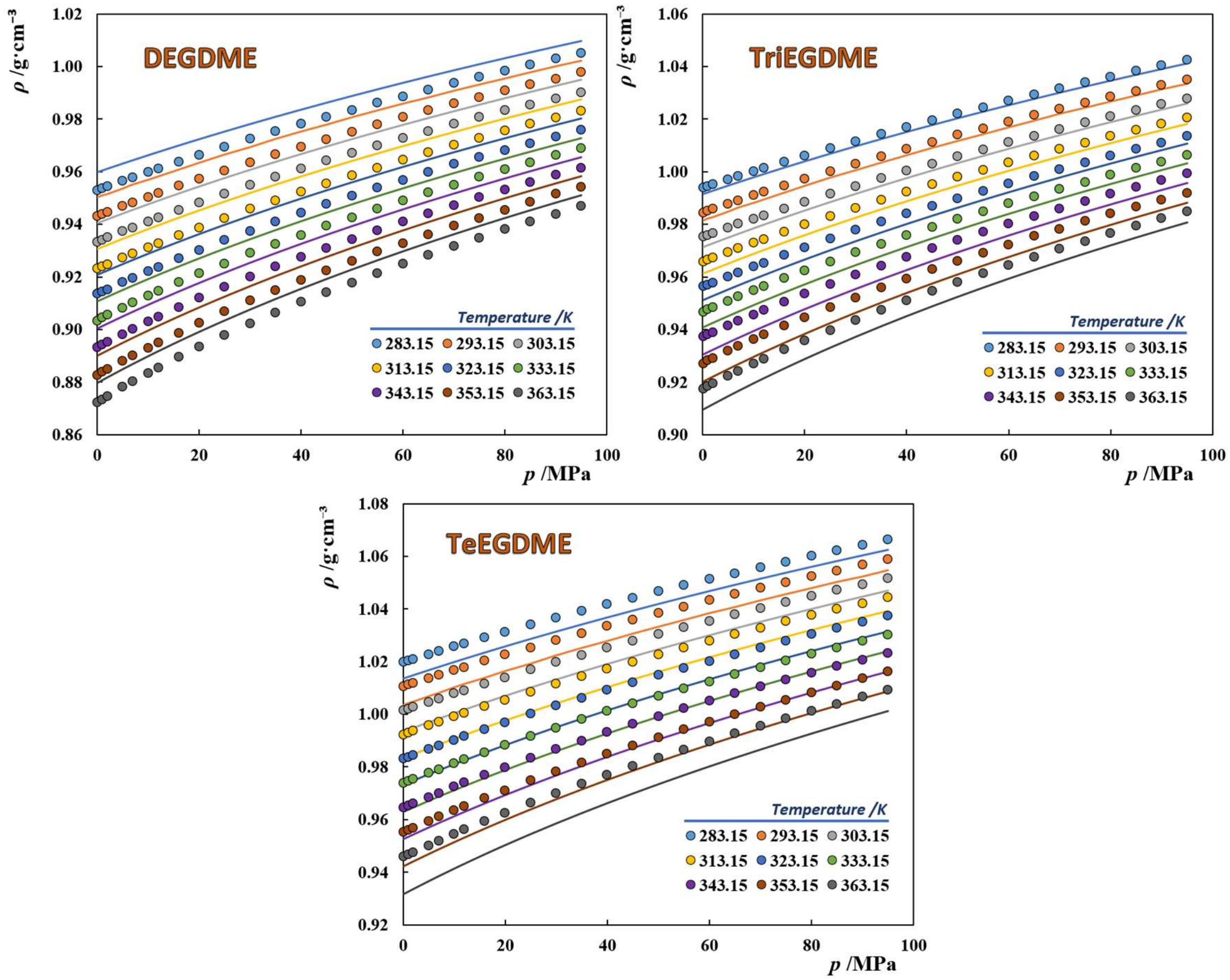

***Figure S25.*** *High-pressure liquid densities of pure glymes. Symbols represent experimental data*[3] *while the solid lines depict the SAFT-γ-Mie results following approach G.*